\documentclass[12pt]{book}
\usepackage{sussp}
\usepackage{graphicx,subfigure}

\def\msun{{\rm\,M_\odot}}
\def\msun{{\rm\,M_\odot}}

\def\spose#1{\hbox to 0pt{#1\hss}}
\def\lta{\mathrel{\spose{\lower 3pt\hbox{$\mathchar"218$}}
     \raise 2.0pt\hbox{$\mathchar"13C$}}}
\def\gta{\mathrel{\spose{\lower 3pt\hbox{$\mathchar"218$}}
     \raise 2.0pt\hbox{$\mathchar"13E$}}}

\def\sech{\mathop{\rm sech}\nolimits}
\def\apj{ApJ}

\def\mnras{MNRAS}
\def\apjs{ApJS}

\makeindex

\begin{document}

\headings{Evolution of galaxies due to self-excitation}
{Evolution of galaxies due to self-excitation}
{Martin D. Weinberg}
{Department of Physics \& Astronomy, University of Massachusetts, Amherst, MA 01003-4525, USA}

\section{Introduction}

These lectures will cover methods for studying the evolution of
galaxies since their formation.  Because the properties of a galaxy
depend on its history, an understanding of galaxy evolution requires
that we understand the dynamical interplay between all components over
10 gigayears.  For example, lopsided ($m=1$) asymmetries are transient
with gigayear time scales, bars may grow slowly or suddenly and, under
circumstances may decay as well.  Recent work shows that stellar
populations depend on asymmetry.

The first part will emphasize n-body simulation methods which minimize
sampling noise.  These techniques are based on harmonic expansions and
scale linearly with the number of bodies, similar to Fourier transform
solutions used in cosmological simulations.  Although fast, until
recently they were only efficiently used for small number of
geometries and background profiles.  I will describe how this
so-called {\em expansion}\index{expansion method} or {\em
  self-consistent field}\index{self-consistent field method} method
can be generalized to treat a wide range of galactic systems with one
or more components.  We will work through a simple but interesting
two-dimensional example relevant for studying bending modes.

These same techniques may be used to study the modes and response of a
galaxy to an arbitrary perturbation.  In particular, I will describe
the modal spectra of stellar systems and role of damped modes which
are generic to stellar systems in interactions and appear to play a
significant role in determining the common structures that we see.  The
general development leads indirectly to guidelines for the number of
particles necessary to adequately represent the gravitational field
such that the modal spectrum is resolvable.  I will then apply these
same excitation to understanding the importance of noise to galaxy
evolution.

\section{N-body simulation using the expansion method}

\subsection{Potential solver overview}

A number of n-body\index{n-body} potential solvers have already been
mentioned in other lectures.  To better understand the motivation for
the development here, I will begin by briefly reviewing and
contrasting their properties.  Many of these have already been
reviewed by Hugh Couchman but I would like to make a general point to
start: the n-body problem of the galactic dynamicist or cosmologist
differs considerably from the n-body problem of the celestial
mechanician or the student of star clusters.  For galactic or CDM
simulations, one really wants a solution to the collisionless
Boltzmann equation\index{collisionless Boltzmann equation} (CBE), not
the an n-body system with finite $N$.  A direct solution of the CBE is
not feasible, so simulate a galaxy by an intrinsically collisional
problem of n-bodies but with parameters that best yield a solution to
the CBE.  In other words, you should consider an n-body simulation in
this application as algorithm for Monte Carlo solution of the CBE.
The $N$ bodies should be considered tracers of the density field that
we simultaneously use to solve for the gravitational potential and
sample the phase-space density.

\subsubsection{Direct summation: the textbook approach}

This truly is the standard n-body problem.  The force law is the exact
pairwise combination of central force interactions; there are $
\left({N\atop2}\right) = {N!\over (N-2)!2!} = {N(N-1)\over 2} $
couplings.  One might use Sverre Aarseth's advanced techniques for
studying star clusters or various special purpose methods to study the
solar system as Tom Quinn and others have reviewed in this volume.

Considered as a solution to the CBE, the density is a distribution of
points and the force from pairwise attraction of all points.  For any
currently practical value of $N$, this system is a poor approximation
to the limit $N\rightarrow\infty$.  Furthermore, the direct problem is
is this very expensive.  Of course, this direct approach is easy to
understand, implement, and with appropriate choice of softening
parameter is useful in some cases.  However in most cases, it makes
sense to take a different approach: interpret the distribution of $N$
points as a sampling of the true distribution.  This motivates tree
and mesh codes among others.

\subsubsection{Tree code}

\begin{figure}[thb]
  \centering
  \subfigure[Tree algorithm]{\includegraphics[width=1.5in]{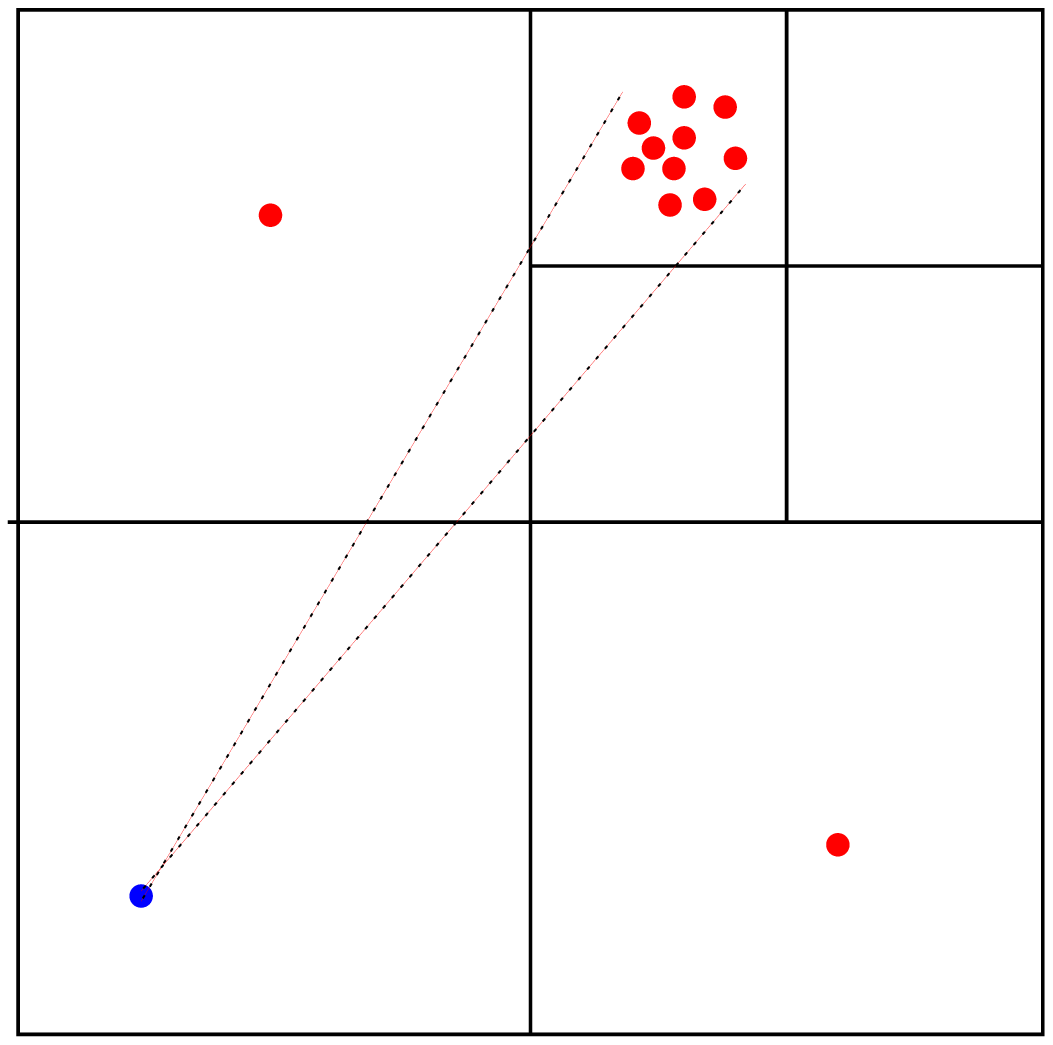}\hspace{10pt}}
  \subfigure[Mesh algorithm]{\hspace{10pt}\includegraphics[width=1.5in]{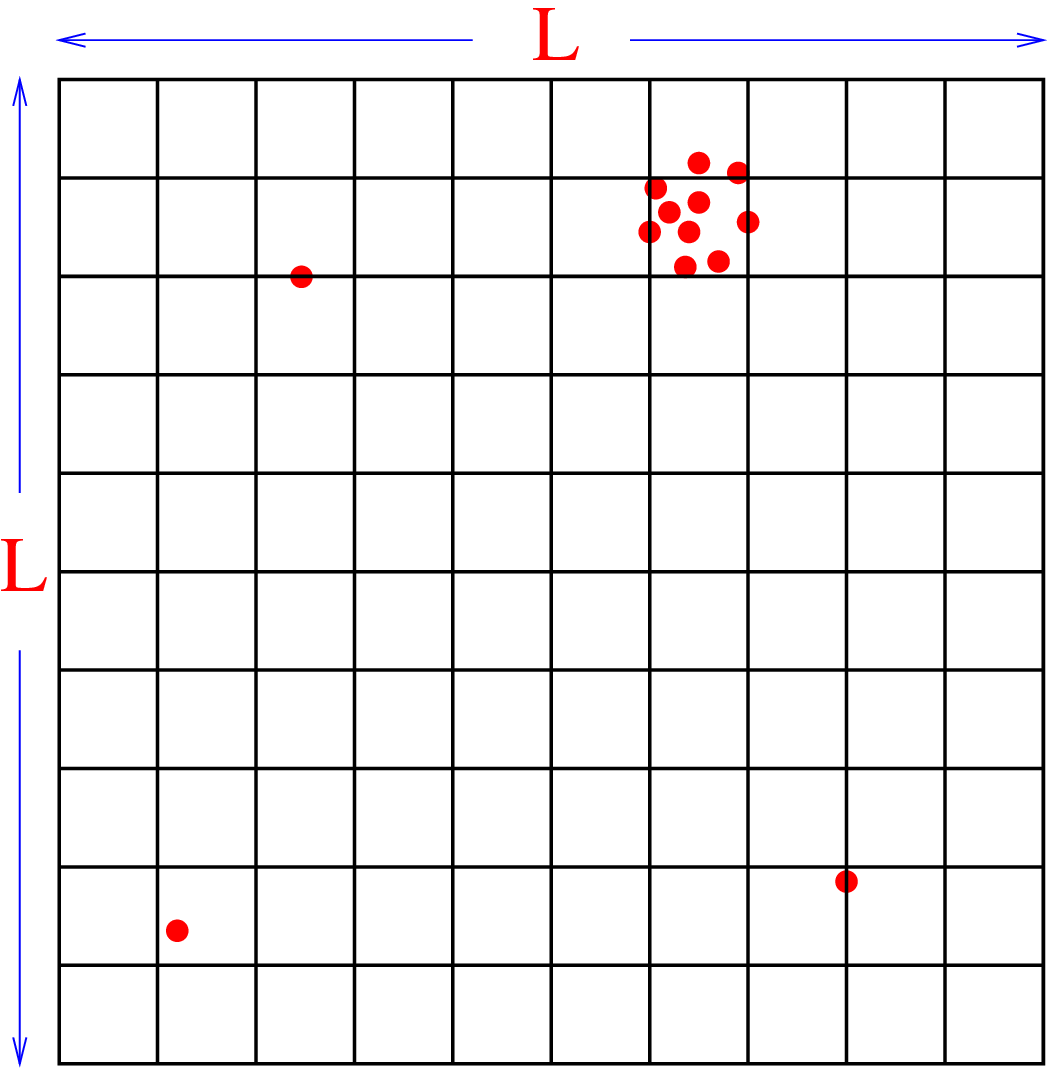}}
  \caption{Construction of the data structure for the tree algorithm in
    two dimensions (a) illustrating the opening angle and (b) the mesh
    algorithm.}
  \label{fig:tree}
\end{figure}

The tree algorithm\index{tree algorithm} makes use of differences in
scales to only do the computational work that will make a difference
to the end result.  The algorithm treats distant groups of particles
as single particles at their centers of mass.  The criterion for
replacing a group by a single particle is whether or not the angular
subtent of that group is smaller than some critical {\em opening
  angle} $\theta_c$.  Figure \ref{fig:tree} shows the recursive
construction that gives the tree code it's name.  This particular tree
is a quad tree although k-d trees and others have been used.  The
force computation only ``opens'' the nodes of the tree if they are
larger than $\theta_c$.  Thinking in terms of multipole expansions,
one is keeping multipoles up to order $l\sim2\pi/\theta_c$; typical
opening angles have $l\gta20$.

\subsubsection{Mesh code}

A mesh code\index{mesh code} is simple in concept.  The steps in the
algorithm are as follows.  First, assign the particle distribution to
bins.  Be aware there are good and bad ways of doing this.  For
example, one may wish distribute the mass of a particle according to a
smoothing kernel rather than using the position and bin boundaries
naively.  Then, represent density as a Fourier series by performing a
discrete Fourier transform by FFT.  Again, one must be very careful
about boundary conditions; see Couchman's paper in this volume and
references therein.  Finally, the gravitational potential follows
directly from Fourier analysis: $ \rho = \sum_{\bf k} c_{\bf
  k}e^{i{\bf k}\cdot{\bf x}} $ and by a simple application of the
Poisson equation, this yields $ \Phi = -\sum_{\bf k} c_{\bf k}
\exp({i{\bf k}\cdot{\bf x})} /4\pi G k^2$.

In short, we are using a mesh to represent the density and exploiting
harmonic properties of the Poisson equation to write down the
gravitational potential.  Note that the particle distribution traces
the mass but an individual particle does not interact with others as a
point mass.

\subsubsection{SPH}

This notion of density representation is explicit in smoothed-particle
hydrodynamics\index{smoothed-particle hydrodynamics} (SPH), a topic
which has also appeared several times in these lectures.  In SPH, the
gas particles must be considered as tracers of the gaseous density,
temperature, and velocity fields.  The hydrodynamic equations are
solved, crudely speaking, by a finite difference solution on an
appropriately smoothed fields determined from the tracers.  One can
show that these algorithms reduce to Euler's equations in the limit of
large $N$. The choice of algorithm and smoothing kernel must be done
with great care but most clearly, the gas particles are not stars or
gas clumps in any physical sense but tracers of field quantities.

\subsubsection{Summary}
All of these but direct summation are examples of {\em density
  estimation}:\index{density estimation} a statistical method for
determining the density distribution function based on a sample of
points.  The algorithms follow the same pattern: (1) Estimate the
density profile of the galaxy based on the $n$ bodies; (2) Exploit
some property of the estimation to efficiently compute the
gravitational potential, and in the case of SPH, other necessary field
quantities; (3) Use the gravitational field to derive the
accelerations, and in the case of SPH, the hydrodynamical equations of
motion.

\subsection{Expansion method}

The expansion method\index{expansion method} is density estimation
using an orthogonal function expansion.  This is a standard technique
in functional approximation and familiar to most readers.  Its
application to solving the Poisson equation is directly analogous to
the grid method.  In the standard grid method, one represents the
density as a Fourier series
\begin{equation}
\rho({\bf r}) = {1\over L^3}
\sum_{l,m,n=-M}^M c_{lmn} e^{i\Delta k(lx + my + nz)} 
\end{equation} 
where $\Delta k=2\pi/L$ and the infinite sum of integer is truncated
at $\pm M$.  Then, by separation of variables, the gravitational
potential is:
\begin{equation}
\Phi({\bf r}) = -{1\over4\pi G (\Delta k)^2}
{\sum^M_{l,m,n=-M}}^\prime c_{lmn} 
{e^{i\Delta k(lx + my + nz)}\over l^2 + m^2 + n^2}.
\end{equation}

There is a way to skip the binning and FFT steps altogether.
We can write the density profile of the $n$ point particles as
\begin{equation}
\rho(x, y, z) = \sum_{i=1}^N \delta(x-x_i)\delta(y-y_i)\delta(z-z_i)
\end{equation}
The coefficient $c_{lmn}$ is integral
\begin{equation}
{1\over L^3}
\int^{L/2}_{-L/2} dx
\int^{L/2}_{-L/2} dy
\int^{L/2}_{-L/2} dz
e^{-i\Delta k(lx + my + nz)}\rho(x, y, z)
\end{equation}
which immediately yields
\begin{equation}
c_{lmn} = {1\over L^3} \sum_{i=1}^N
e^{-i2\pi l x_i/L} e^{-i2\pi m y_i/L} e^{-i2\pi n z_i/L}
\end{equation}
and we are done!  From these coefficients, we have the potential and
force fields.  This may be less efficient than an FFT scheme in some
cases and suboptimal density estimation because the lack of smoothing
may increase the variance, but it is applicable to non-Cartesian
geometries for which no FFT exists as we will see below.

\subsection{General theory for gridless expansion}
\label{sec:gridless}

We tend to take for granted special properties of sines and cosines in
solving the Poisson equation.  However, most of the special properties
are due to the equation not the rectangular coordinate system.  In
particular, the Poisson equation is separable in all conic coordinate
systems (e.g. \cite{Morse.Feshbach:53}). Each of separated equation
takes the Sturm-Liouville\index{Sturm-Liouville equation} (SL) form:
\begin{equation}
{d\over dx}\left[p(x){d\Phi(x)\over dx}\right]
- q(x)\Phi(x) = \lambda w(x)\Phi(x)
\end{equation}
where $p(x), q(x), w(x)$ are real and $w(x)$ is non-negative.  The
eigenfunctions of this equation are {\em orthogonal} and {\em
  complete}!  The implications of this is the existence of pairs of
functions, one representing the density and one the potential, that
are mutually orthogonal and together can be arranged to satisfy the
Poisson equation.  Such a set of pairs is called {\em biorthogonal}.
\index{biorthogonal basis} Just as in the case of rectangular
coordinates, the particle distribution can be used to determine the
coefficients for a biorthogonal basis set and the coefficients yield a
potential and force field.

\subsubsection{Pedagogical example: semi-infinite slab}
\label{sec:slabex}

Here, we will develop a simple but non-trivial example of a
biorthogonal basis.  Our system is a slab of stars, infinite in $x$
and $y$ directions but finite in $z$; that is, $\rho=0$ for
$|z|>L$.\index{semi-infinite slab} Since the coordinates are
Cartesian, the eigenfunctions of the the Laplacian (the SL equation)
are sines and cosines again and we do not have to construct a explicit
solution.  The subtlety in the solution is the proper implementation
of the boundary conditions.

Proceeding, we know that we should find a biorthogonal basis of
density potential-density pairs, $p_\mu, d_\mu$, with a scalar product
\begin{equation}
(p_\mu, d_\nu) = -\int dx dy dz p_\mu^\ast d_\nu = \delta_{\mu\nu}
\end{equation}
such that $\nabla^2 p_\mu = d_\mu$.  Inside the slab, solutions are
sines and cosines in all directions.  However, outside slab, the
vertical wave function must satisfy the Laplace equation
\begin{equation}
{d^2\Psi\over dz^2} -k_x^2\Psi  = 0
\end{equation}
which has the solution
\begin{equation}
{d\Psi\over dz} \propto \cases{ 
  e^{-k_xz}  & $z\ge L$ \cr
  e^{k_xz} & $z\le -L$ \cr
  }
\end{equation}
where $k_x$ is the wave vector in the horizontal direction.  The
Laplacian is self-adjoint with these boundary conditions. Therefore,
the resulting eigenvalue problem is of Sturm-Liouville type whose
eigenfunctions are a complete set.

Taking the form $\Psi = A\cos(kz+\alpha)$ results in the following
requirements on $k$: $\alpha=m\pi/2$ and
\begin{equation}
\cases{
  \tan(kL)=k_x/k & $m$ even, \cr
  \cot(kL)=-k_x/k & $m$ odd. \cr
  }
\end{equation}
Let $k^e_{\ast n}$ and $k^o_{\ast n}$ be the solutions of these two
relations where $k^e_{\ast n}\in[n\pi,n\pi+\pi/2]$ and $k^o_{\ast
  n}\in[n\pi+\pi/2,(n+1)\pi]$ .  The normalized eigenfunctions are
$\Psi^e_n = A^e_n \cos(k^e_{\ast n}z)$ and $\Psi^o_n = A^o_n
\sin(k^o_{\ast n}z)$ with normalization constants $A^e_n$ and $A^o_n$.
Finally, putting all of this together, the biorthogonal pairs can be
defined as
\begin{equation}
p_{\mu\,{\bf k}} = {[k_\ast^2 + k^2]^{-1/2} \over 2\pi }\ 
\Psi_\mu(z) e^{i{\bf k}\cdot{\bf R}},  \qquad
d_{\mu\,{\bf k}} = {[k_\ast^2 + k^2]^{1/2} \over 2\pi }\ 
\Psi_\mu(z) e^{i{\bf k}\cdot{\bf R}}
\end{equation}
where ${\bf k}$ and ${\bf R}$ are vectors in the $x$-$y$ plane and
$\Psi_n$ and $k_\ast$ denote both the even and odd varieties.  The
orthogonality relationship is
\begin{equation}
-\int d^3x\, p_{\mu\,{\bf k}}^\ast d_{\nu\,{\bf k}^\prime} = \delta_{\mu\nu}
\delta({\bf k}-{\bf k}^\prime).
\end{equation}

The application to an n-body simulation requires two ${\cal O}(N)$
steps:
\begin{enumerate}
\item We obtain the coefficients by summing the basis functions
  over the $N$ particles:
  $c_{\mu\,{\bf k}} = \sum_{i=0}^N m_i p_{\mu\,{\bf k}}({\bf R}_i, z_i)$
  where ${\bf k}=(k_x, k_y)$ is the in-plane wave vector now
  generalized to remove in identification of ${\hat k}={\hat x}$ and
  ${\bf R}=(x, y)$.
\item We compute the force force by gradient of potential:
  $F({\bf r}) = -{\partial\over\partial{\bf r}} 
  \sum_{\bf k} d{\bf k} \sum_\mu p_{\mu\,{\bf k}}({\bf R}, z).$
  Because the slab is unbounded in the horizontal direction the values
  of ${\bf k}$ are continuous and therefore, construction of the
  potential requires an integral over ${\bf k}$.  This is indicated as
  a discrete sum over the volume in ${\bf k}$ space in the expression
  for $F({\bf r})$.
\end{enumerate}

A few short words about error analysis for this scheme.  Nearly all
results follow from the identification of this algorithm as a specific
case of linear least squares \cite{Dahlquist.Bjork:74}.  For our
purposes, it is interesting to note the coefficient determination in
the expansion method is, therefore, unbiased: $\hbox{E}\{c_\mu\} =
{\bar c}_\mu$.  This means that if one performs a large number of
Monte Carlo realizations, the expectation value of the coefficients
from this ensemble will be the true values.  One can derive formal
error estimates for method, following the approach outlined in many
standard probability and statistics texts.  In this case we find that
\begin{equation}
\hbox{Var} \propto {\mu_{max}\over N}
\label{eq:variance}
\end{equation}
where $\mu_{max}$ is the maximum order in the expansion series and $N$
is the number of sample points.  This is broadly consistent with
expectations: the variance in a Monte Carlo estimate scales as $1/N$
and each independent parameter contributes to this variance.  More
informative analyses are possible.  In particular, it is
straightforward to compute the variance of the coefficients (or the
entire covariance matrix) and estimate the the {\em signal to noise}
ratio for each coefficient.  Then, one may truncate series when
information content becomes small, or at the very least, use this
information to inform future choices of $\mu_{max}$ (see Hall, 1981
\nocite{Hall:81} for general discussion in the density estimation
context).

\subsubsection{Example: spherical system}

The recurring slab example in this presentation is intended to give
you a complete example which illustrates most aspects of the method,
rather than be of use for a realistic astronomical scenario.
Nonetheless, it is easy to implement and coupled with the analytic
treatment in \S\ref{sec:slabdisp} is useful for exploring the effects
of particle number (more on this below).

Astronomically useful geometries include the spherical, polar and
cylindrical bases, although as mentioned above, this approach can be
applied to any conic coordinate system.  For example, the Poisson
equation separates in spherical coordinates and each equation yields
an independently orthogonal basis: (1) trigonometric functions in the
azimuthal direction, $e^{im\phi}$; (2) associated Legendre polynomials
in latitudinal direction, $P_l^m(\cos\theta)$; and (3) Bessel
functions in the radial direction, $q_{nl} J_{l+1/2}(\alpha_n r/R)$.
The first two bases combine to form the spherical harmonics,
$Y_{lm}(\theta,\phi)$.  The $\alpha_n$ follow from defining physical
boundary conditions that the distribution vanishes outside of some
radius $R$ and $q_{nl}$ is a normalization factor.  This bit of
potential theory should be familiar to readers who have studied
mathematical methods of physics or engineering.
  
For $N$ sampled particles at position ${\bf r}_i$, the gravitational
potential is then
\begin{equation}
  V(r, \theta, \phi) = \sum_{l=0}^{l_{max}} \sum_{m=-l}^l
 \sum_{\mu=1}^{\mu_{max}}  Y_{lm}(\theta_i,\phi_i) q_{\mu l}
 J_{l+1/2}(\alpha_\mu r/R),
\end{equation}
where the expansion coefficients are
\begin{equation}
  c_{\mu lm} = -4\pi G \sum_{i=1}^N Y_{lm}(\theta_i,\phi_i) q_{\mu l}
  J_{l+1/2}(\alpha_\mu r_i/R).
\end{equation}
This set is is easy to describe but the basis functions look nothing
like a galaxy.  Therefore, one requires many terms to represent the
underlying profile and any deviations.  Because the variance increases
with $\mu_{max}$ (cf. eq. \ref{eq:variance}), such a basis is
inefficient.

\subsection{Basis Sets}

There is an obvious way around this problem.  Nothing requires us to
use the Bessel function basis directly and we can construct new bases
by taking weighted sums to make lowest order member have any desired
shape.  

This method is nicely described in
\cite{Clutton-Brock:72,Clutton-Brock:73} by Clutton-Brock who shows
that a suitably chosen coordinate transformation followed by an
orthogonality requirement, leads to a recursion relation for a set of
functions whose lowest order members do look like a galaxy.  He
describes two sets in each of these papers, a spherical set whose
first member is proportional to a Plummer model and two-dimensional
polar set whose first member is similar to a Toomre disk.  At nearly
the same time, Kalnajs described two-dimensional set appropriate for
studying spiral modes \cite{Kalnajs:76,Kalnajs:77}.  More recently,
Hernquist \& Ostriker \cite{Hernquist.Ostriker:92} used
Clutton-Brock's construction to derive a basis whose lowest-order
member is the Hernquist profile \cite{Hernquist:90}.

The lack of choice in basis functions in all but a few cases, however,
seems to have limited the utility of the expansion approach.  But,
there is really no need for analytic bases (or those constructed from
an analytic recursion relation) are not necessary.  Saha
\cite{Saha:93} advocates constructing bases by direct Gram-Schmidt
orthogonalization beginning with any set of convenient functions.
Recall from \S\ref{sec:gridless} that the original motivation for
using eigenfunctions of the Laplacian is that these are solutions to
the Sturm-Liouville equation\index{Sturm-Liouville equation} and
therefore orthogonal and complete.  The SL equation has many useful
properties and recently these have lead to very efficient methods of
numerical solution \cite{Pruess.Fulton:93}.  By numerical solution, we
can construct spherical basis sets with any desired underlying profile
and three-dimensional disk basis sets close to a desired underlying
profile \cite{Weinberg:99}.  The next section describes the method.

\subsection{Empirical bases}

The spherical case is straightforward and illustrates the general
procedure.  We still expand in spherical harmonics and only need
to treat the radial part of the Poisson equation:
\begin{equation}
{1\over r^2}{d\over dr} r^2 {d\Phi(r)\over dr} + {l(l+1)\over
  r^2}\Phi(r) = 4\pi G\lambda \rho(r).
\label{eq:slrad}
\end{equation}
The most important point is to search for solutions of the form
$\Phi(r) = \Psi_o(r) u(r), \rho(r) = \rho_o(r) u(r)$n where
$\Psi_o(r)$ and $\rho_o(r)$ are {\em conditioning} functions.

Note that if we were to choose our conditioning functions so that
$\nabla^2\Psi_o(r) = 4\pi G\rho_o(r)$, the lowest order basis function
will be a constant, $u(r)=\hbox{constant}$, with unit eigenvalue
$\lambda=1$.  In words, by choosing $\Psi_o$ appropriately, we have
achieved the goal of a basis whose lowest order member can be chosen
to match the underlying profile and, furthermore, the entire basis will
be orthogonal and complete.

\begin{figure}[thb]
  \mbox{
    \includegraphics[width=2.1in]{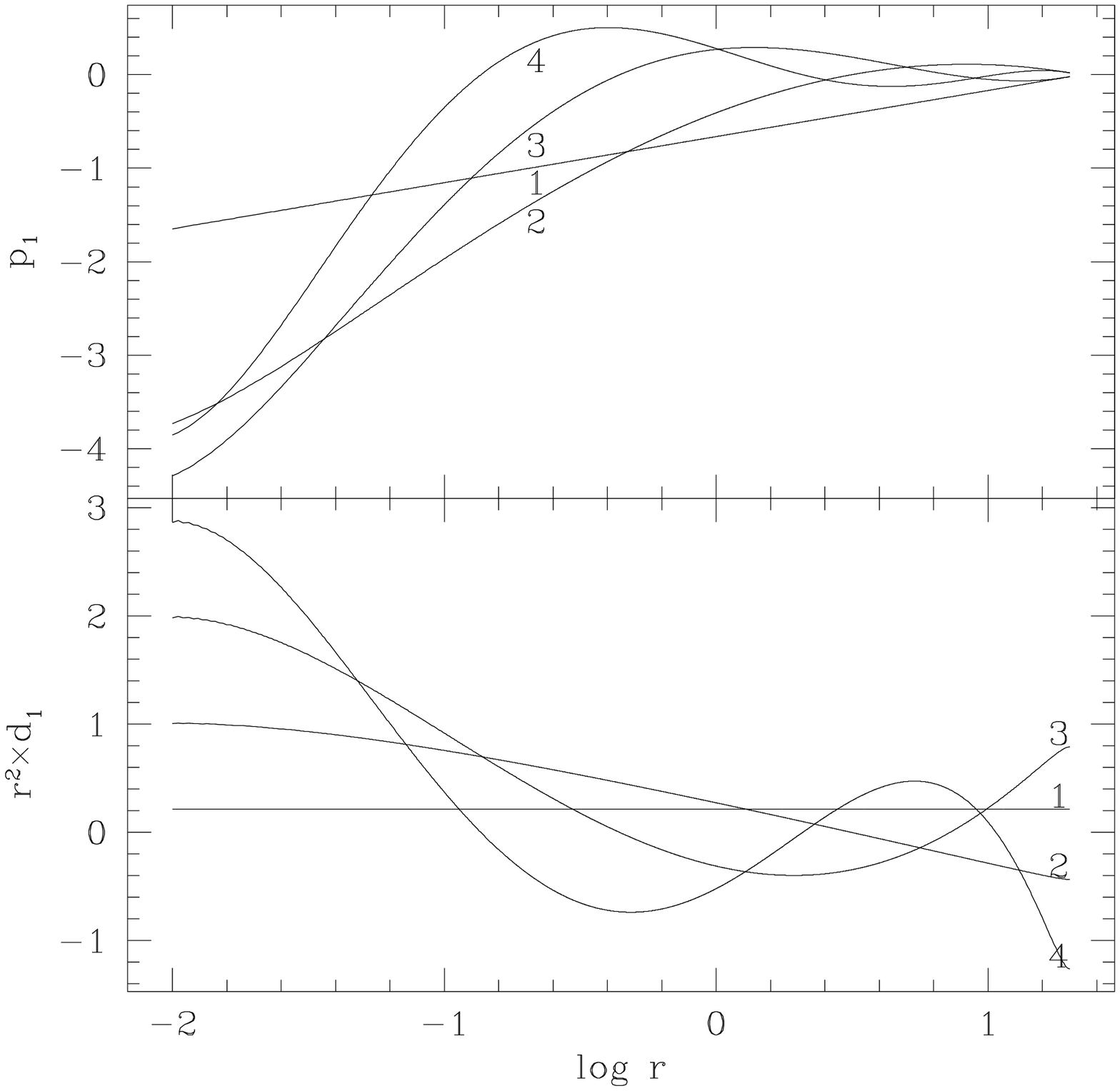}
    \includegraphics[width=2.1in]{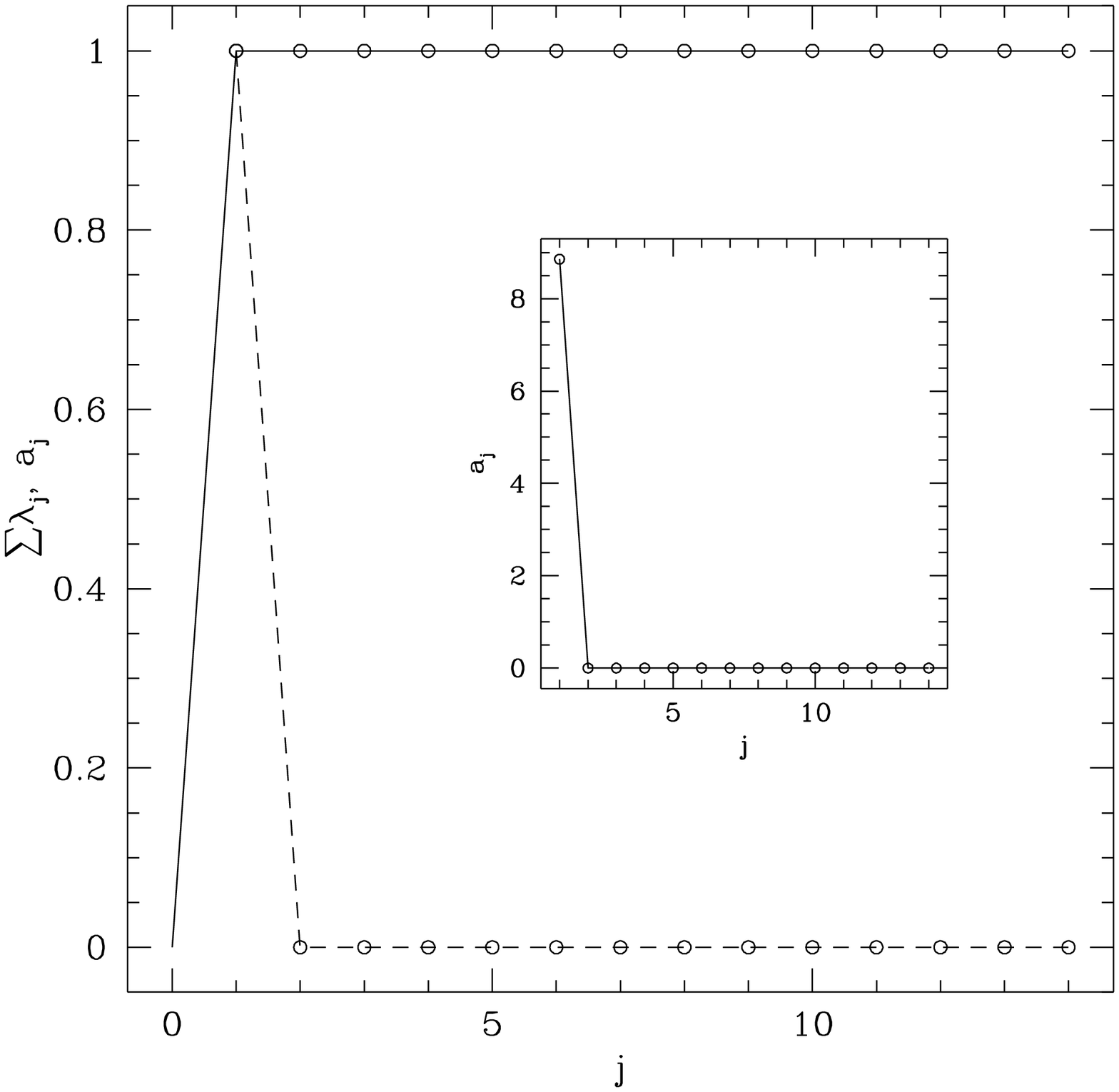}
    \includegraphics[width=2.1in]{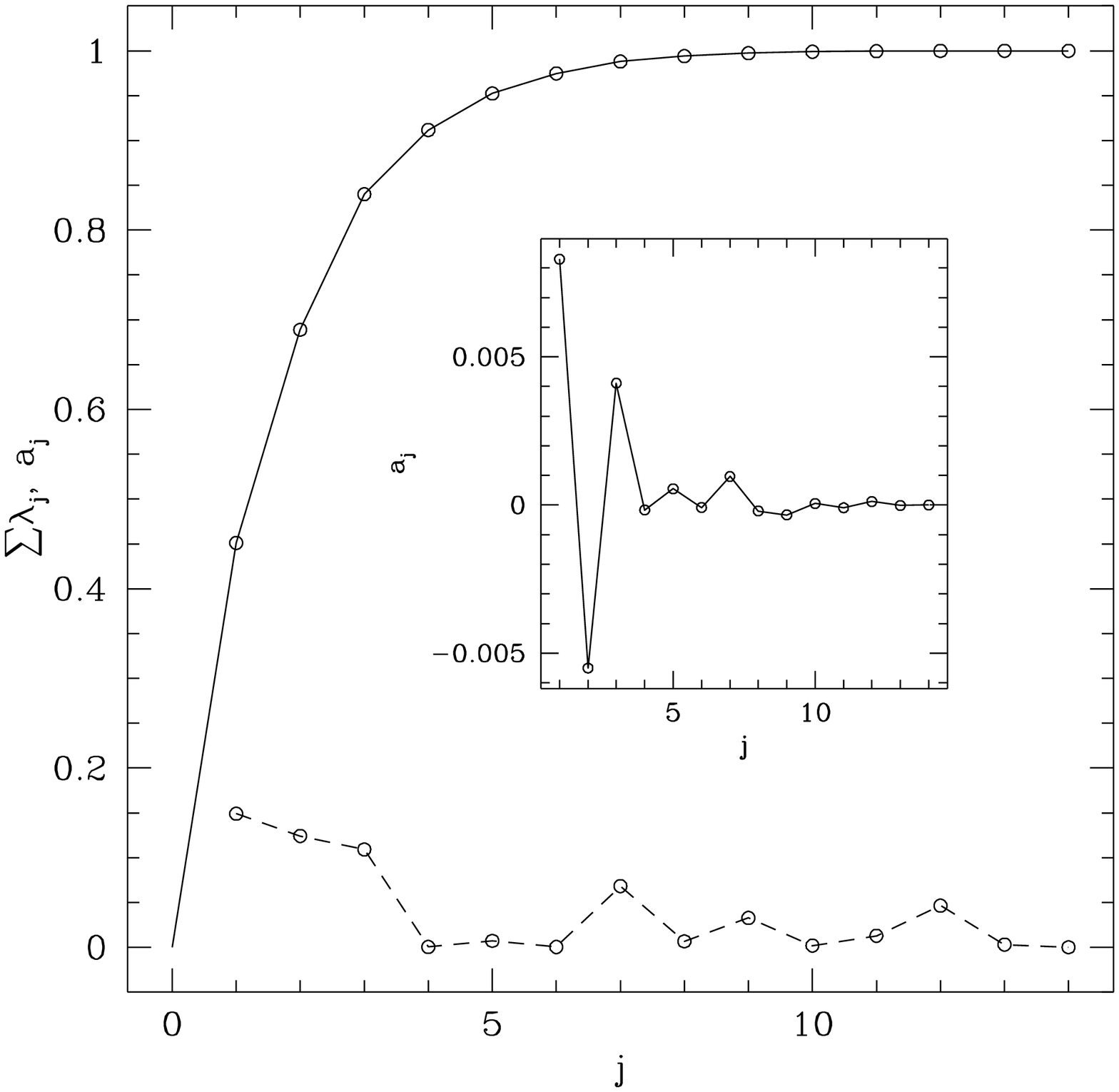} 
    }
  \caption{Left: Basis derived assuming the singular isothermal sphere
    profile as conditioning functions.  The upper (lower) panel shows
    the potential (density) members for harmonic $l=0$.  The density
    members are premultiplied by $r^2$ to suppress the dynamical
    range. Convergence of the coefficients for a Monte Carlo
    realization of the underlying profile for $l=0$ (center) and $l=2$
    (right).  The solid line (dashed line) shows the cumulative
    explained variance (values of the coefficients).}
  \label{fig:converge}
\end{figure}

Figure \ref{fig:converge} shows an example conditioned to the singular
isothermal sphere, a case that would be challenging for other the
standard bases (and other potential solvers).  Note that the lowest
order members have potential and density proportional to $\ln r$ and
$r^{-2}$.  Each successive member has an additional radial node.

Figure \ref{fig:converge} illustrates the advantage of the basis by
illustrating the convergence of the coefficients for a Monte Carlo
simulation of $N=10^5$ particles.  The $l=0$ plot shows that all of
the {\em variance} in the distribution is described by the lowest
order basis function as expected by design.  The $l=2$ case is noise;
the plot shows that nearly all of the variance is described by with
$j\lta8$.

A main deficiency of the expansion method has been the lack of
suitable bases for simulating a galactic disk with non-zero scale
height.  This can also be accomplished by direct solution of the
Sturm-Liouville equation but with an additional complication: we can
only use the conditioning trick in one dimension.  For the cylindrical
disk, the separable equations give us trigonometric functions in both
the azimuthal and vertical dimensions.  A related approach has been
described by Robijn \& Earn \cite{Robijn.Earn:96} but users must take
care to apply appropriate boundary conditions.  We now have a choice,
we can condition in $z$ or $R$.  The other dimension can be
orthogonalized ex post facto to provide a good match to the underlying
distribution using an empirical orthogonal function
analysis\index{empirical orthogonal functions} (also known as
principal component analysis).

Explicitly, the Laplace equation separates in cylindrical coordinates
using $\Psi({\bf r}) = R(r)Z(z)\Theta(\theta)$ as follows:
\begin{eqnarray}
  {1\over r}{d\over dr}r{d\over dr} R(r) -\left(k^2 +
    {m^2\over r^2}\right) R(r) &=& 0 \nonumber \\
  {d^2\over dz^2}Z(z) + k^2 Z(z) &=& 0 \nonumber \\
  {d^2\over d\theta^2}\Theta(\theta) + m^2 \Theta(\theta) &=& 0
\label{eq:laplace}
\end{eqnarray}
As in the spherical case, let us assume solutions of the form $\Psi(r,
z, \theta) = \Psi_o(r) u(r) Z(z) \Theta(\theta)$ and $\rho(r, z,
\theta) = \rho_o(r) u(r) Z(z) \Theta(\theta)$ with radial conditioning
functions.  The Poisson equation becomes
\begin{equation}
  {1\over r}{d\over dr}r{d\over dr} \Psi_o(r) u(r) -\left(k^2 +
    {m^2\over r^2}\right) R(r) = 4\pi G \lambda \rho_o(r) u(r)
\end{equation}
together with second two of equation (\ref{eq:laplace}) above, where
$\lambda$ is an unknown constant.
In SL form, this is:
\begin{eqnarray}
  {d\over dr}\left[ r\Psi_o^2(r) {d u(r)\over dr} \right] - 
  \left[ k^2 \Psi_o(r) + {m^2\over r^2}\Psi_o(r) -
    \nabla_r^2\Psi_o(r)\right] r \Psi_o(r) u(r) = \nonumber \\
  \qquad 4\pi G \lambda r\Psi_o(r) \rho_o(r) u(r)
\end{eqnarray}

Now, use standard SLE solver to table the eigenfunctions.  These
coefficient functions now provide the input to the standard packaged
SLE solvers either in tabular or subroutine form.  The orthogonality
condition for this case is
\begin{equation}
  \label{eq:ortho}
  -4\pi G \int^\infty_0 dr\, r\, \Psi_o(r)\rho_o(r) u(r)^2 = -4\pi
  G\int^\infty_0 dr\, r \,\Psi \rho = 1.
\end{equation}
The functions $\Psi(r, z, \theta)$ and $\rho(r, z, \theta)$ are
potential-density pairs.  Just as for the spherical case, the lowest
eigenvalue is unity and the corresponding eigenfunction $u(r)$ is a
constant function if $\Psi_o$ and $\rho_o$ solve Poisson equation.
Again $\Psi_o$ and $\rho_o$ need not solve the Poisson equation, but
the conditioning functions must obey appropriate boundary conditions
at the center and at the edge.  This is especially appropriate for
this cylindrical case where equilibria solutions for three-dimensional
disks are not convenient.

\begin{figure}
  \mbox{
    \includegraphics[width=2.75in]{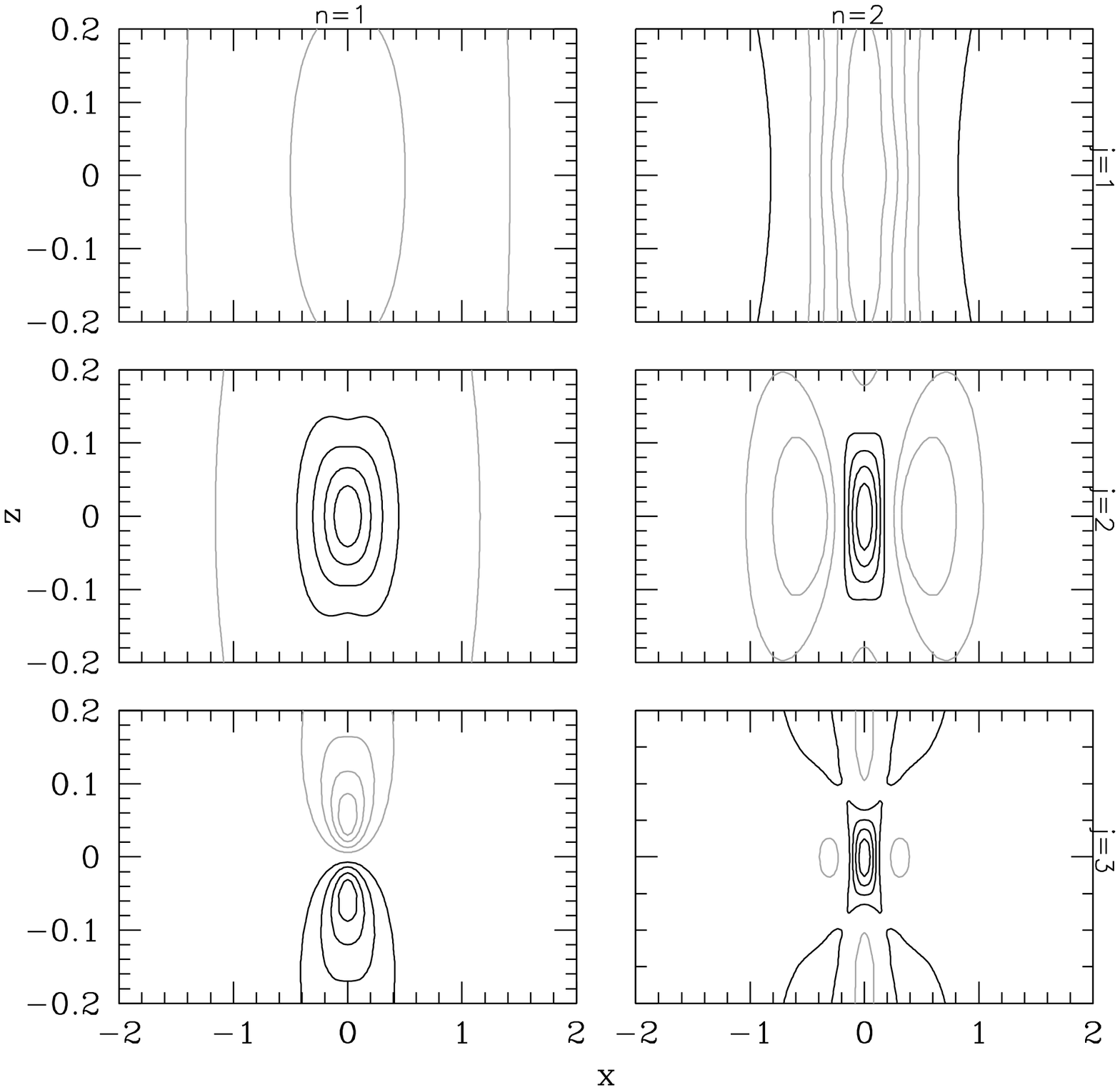} 
    \includegraphics[width=2.75in]{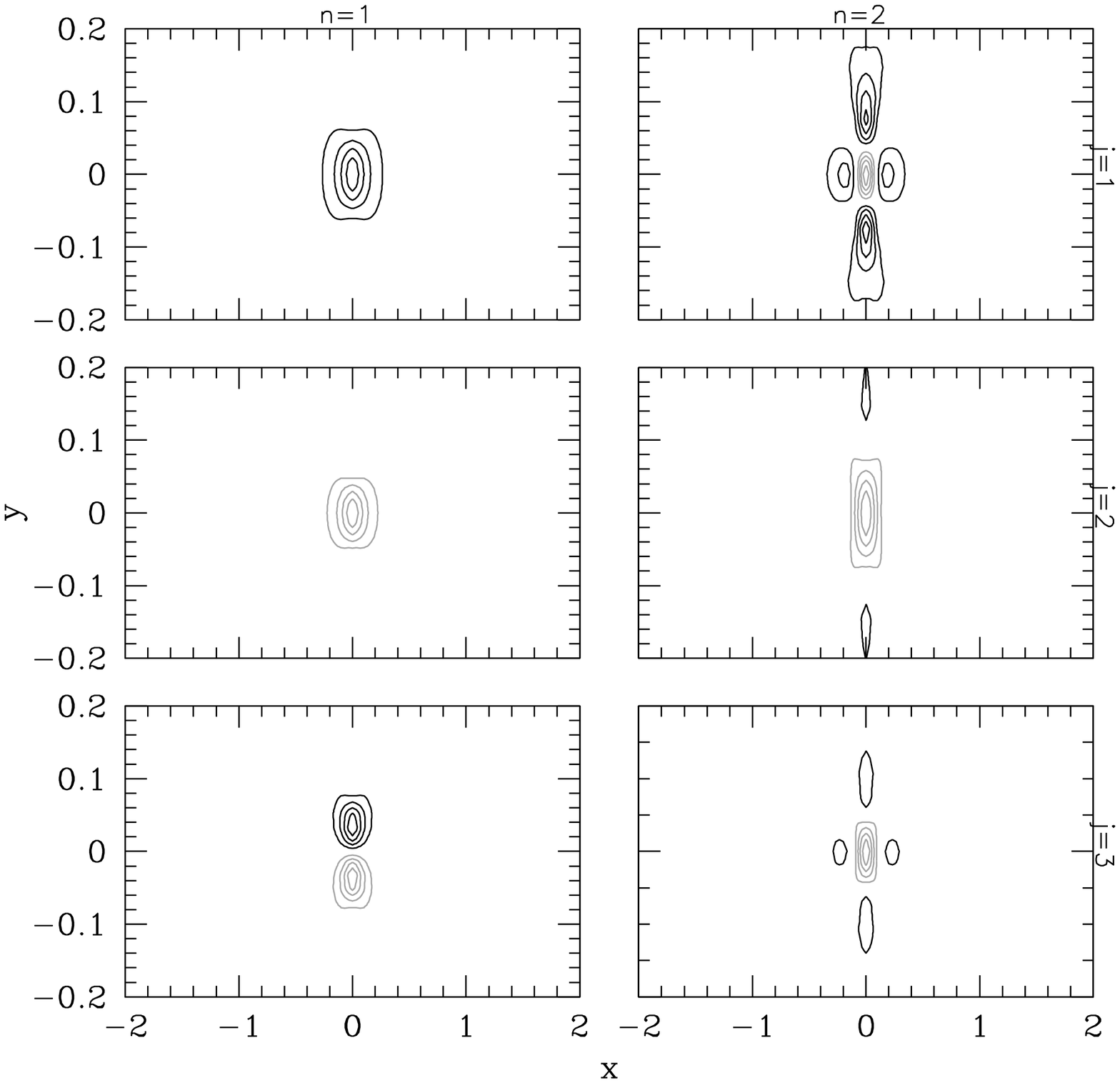} 
    }
  \caption{Cylindrical basis set conditioned by an exponential radial
    density profile and sech-squared vertical profile for $m=0$.  The
    potential (density) members are shown on the left (right) labeled
    by radial and vertical orders.  Positive (negative) isovalues are
    shown as black (gray). }
  \label{fig:cyldisk}
\end{figure}

Figure \ref{fig:cyldisk} shows the basis set for the SL method
conditioned by an exponential radial density profile and $\sech^2$
vertical profile.  The steps in the construction were as follows:
\begin{enumerate}
\item The radial SL equation is solved numerically with conditioning
  functions $\Psi_o(R)\propto(1+(R/a)^2)^{-1}$ and
  $\rho_o\propto\exp{-R/a}$.  The vertical functions are sines and
  cosines with vacuum boundary conditions.
\item Linear combinations of the resulting eigenfunctions are found
  using an empirical orthogonal function analysis to find the best
  description of the $\Psi_o(R)\sech^2(z/h)$ in the least squares
  sense.
\item The resulting basis functions are tabulated and interpolated as
  needed.  Note that the basis can be chosen to have definite parity
  which optimizes table storage.
\end{enumerate}

\subsection{What good is all of this?}

So far, we have explored a general approach for representing the
gravitational potential for an ensemble of particles using particular
harmonic bases.  These bases can be derived in any coordinate system
in which the Poisson equation is separable; at the very least, this
includes all conic coordinate systems.  Other advantages include:
\begin{itemize}
\item This potential solver is {\em fast}: it is ${\cal O}(N)$ with
  small coefficient.  Recall that the most popular approaches: tree
  and grid codes are ${\cal O}(N\ln N)$.  Direct summation is ${\cal
    O}(N^2)$.   For large $N$, this method has optimal scaling.
\item Each term in the expansion resolves successively smaller
  structure. By truncating the series at the minimum resolution of
  interest or when the coefficients have low S/N, the high-frequency
  fluctuations are filtered out.  This approach results in a
  relatively low-noise simulation; the high-frequency part of the
  noise spectrum dominates the particle noise in the standard
  potential solvers.
\item Note that all of the dynamical information in a simulation is
  represented by the expansion coefficients.  In other words, the
  expansion coefficients significantly compress the structural
  information in the simulation.  If the dynamical content of the
  density and potential fields is the goal, one does not need to keep
  entire phase space, only the coefficients.  Similarly, velocity
  fields may be represented by a similar expansion (e.g.
  \cite{Saha:93}).
\item One is not restricted to individual components or single bases
  and can assign parts of phase space to separate bases depending upon
  its geometry or history.  This is precisely what one needs to study
  a disk embedded in a spheroid and halo.
\end{itemize}

There is no one method for solving the Poisson equation in a
simulation.  The major disadvantages of the expansion approach is its
lack of spatial adaptivity.  It efficiently resolves non-axisymmetric
features and disturbances as long as the galaxy does not change its
structure rapidly.  The approach is not highly adaptive and would not
be good for equal mass merger, for example.  Similarly, these schemes
(like most efficient algorithms) do not strictly conserve momentum.
In the limit of a large number of bodies, the expansion center is
arbitrary because the distribution can be represented in an origin
independent way regardless of the expansion center.  For a smaller
number of bodies, the number of available high-signal-to-noise ratio
coefficients is too small to permit resolving the expansion about an
arbitrary origin.  The offset of the origin allowed for a given error
bound decreases for with particle number.  This demands that an
efficient implementation of the expansion-based Poisson solver
recenter the particle distribution.

These advantages, properties and limitations motivate a set of ideal
applications:
\begin{enumerate}
\item {\em Simulating a multicomponent galaxy.}\ A feature of n-body
  simulation of galaxies is the disparate length scales of the disk,
  bulge and halo.  This is not a problem for the expansion.  We can
  pick a separate basis tailored to each component and determine the
  total gravitational field from their sum.
\item {\em Long-term evolution.}\ For a fixed number of particles,
  Poisson fluctuations and the simulation's self-gravitating response
  to those fluctuations limit the length of time ${\cal T}$ that the
  evolution remains a good approximation to collisionless Boltzmann
  equation.  For too few particles, the fluctuations can be so large
  that the angular momentum and energy of a particle orbit can drift
  or {\em diffuse} significantly over a single orbital time.  Because
  the expansion method filters the high-frequency noise by
  construction, this is likely to give the largest value of ${\cal T}$
  in most cases.
\item {\em Weak, cumulative and tidal interactions.}\ Similarly, this
  method is ideal for studying the response of a simulated galaxy to
  external global distortions.  The scale sensitivity can be
  manipulated to efficiently represent the scales of interest and no
  others.  Of course, limiting the resolution a priori is not always
  the best policy and this strategy must be motivated by a prior study
  with weaker constraints.
\item {\em Stability.}\ This Poisson solver is ideally suited to
  studying global stability.  A time-series analysis of the
  coefficients can empirically yield both the growth rate and shape of
  the unstable mode.
\end{enumerate}

\section{A numerical method for perturbation theory}

N-body simulation is not the only use for this special Poisson-solving
biorthogonal expansion.  We can exploit the completeness property to
transform a linearized solution of the collisionless Boltzmann equation
to a system of linear equations.  This has been given the moniker {\em
  matrix method}\index{matrix method} by dynamicists but is a standard
approach to solving partial differential equations
\cite{Courant.Hilbert:53}.  By using the same expansion for both an
analytic linear solution and an n-body simulation, we explore a
particular problem both ways and even apply the two together in
various hybrid ways to further increase the dynamic range or time scale
${\cal T}$.  I will sketch the development in the next section and
follow this with a simple but complete example based on the slab
model.

\subsection{Introduction}

The response of our stellar galaxy to any distortion is mathematically
described by the simultaneously solution of the collisionless
Boltzmann and Poisson equations:
\begin{eqnarray}
  {\partial f\over\partial t} + 
  {\partial H\over\partial {\bf p}}\cdot{\partial f\over\partial {\bf x}} -
  {\partial H\over\partial {\bf x}}\cdot{\partial f\over\partial {\bf p}} 
  &=& 0, \label{eq:boltz} \\ \noalign{\vspace{10pt}}
  \nabla \Phi({\bf x}) = 4\pi G\rho({\bf x}). && \label{eq:pois}
\end{eqnarray}
The steps in the solution are as follows.  First we linearize equation
(\ref{eq:boltz}) and note that equation (\ref{eq:pois}) is already
linear.  We then separate the partial equations in their natural
bases.  In general, the two equations separate different bases and
this presents a technical problem but not an insurmountable one.  The
Cartesian coordinate system is the exception: the bases are the same.

For a spherical stellar distribution, the biorthogonal
potential-density pairs take the following form: $\Phi({\bf r}) =
\sum_{lm} \sum_i a^{lm}_i Y_{lm}(\theta, \phi) \Phi^{lm}_i(r)$ with an
analogous expression for $\rho({\bf r})$.  The two partial
differential equations are then transformed to Fourier space using
these bases to yield a set of algebraic equations.  To do this, we
note that orbits are quasi-periodic in regular potential.  If all
conserved quantities exist then by the averaging principle
\cite{Arnold:78}, we can represent any phase-space quantity by the
following expansion in action and angles:
\begin{equation}
f({\bf p}, {\bf x}) = \sum_{\bf l} f_{\bf l}({\bf
  I})\exp(i{\bf l}\cdot{\bf w}).
\end{equation}
If the gravitational potential admits chaotic orbits, this approach
does not apply strictly.  If the Lyapunov exponents are small,
quasi-periodicity should still be a good approximation.  With these
tools and conditions, we begin by linearizing the collisionless
Boltzmann equation (eq. \ref{eq:boltz}).  After expressing all
phase-space variables in actions and angles, a Fourier transform in
angles followed by a Laplace transform in times yields the solution
\begin{equation}
  {\hat f}_{1\,{\bf l}}({\bf I}) = {i{\bf l}\cdot{\partial f_o/\partial
      {\bf I}} \over s + i{\bf l}\cdot{\bf\Omega}} {\hat H}_{1\,{\bf
      l}}
  \label{eq:fpert}
\end{equation}
where the hat denotes a Laplace transformed quantity and the subscript
${\bf l}$ denotes an action-angle transform.  Finally, we can
integrate equation (\ref{eq:fpert}) over ${\bf v}$ to get $
{\hat\rho}_1({\bf I}, {\bf w})$.  We have not included the
simultaneous solution of the Poisson equation but at this point, we
tie the two together by expanding both ${\hat\rho}_1({\bf I}, {\bf
  w})$ and the perturbing potential in the biorthogonal series.
Explicitly, we can determine the scalar product of the potential
component of the pair with the velocity integral of equation
(\ref{eq:fpert}):
\begin{equation}
  \int dr r^2\Phi^{lm}_i(r) {\hat\rho}_1({\bf I}, {\bf w}) =
  \int dr r^2\Phi^{lm}_i(r) \int d^3v {i{\bf l}\cdot{\partial f_o/\partial
      {\bf I}} \over s + i{\bf l}\cdot{\bf\Omega}} {\hat H}_{1\,{\bf
      l}}.
\end{equation}
The left-hand side is density expansion coefficients ${\bf a}$.  The
right-hand side may be written as the action of matrix on the vector
of coefficients describing the perturbing potential ${\bf b}$.  The
matrix ${\cal R}$ depends on the underlying unperturbed distribution
function and the Laplace expansion frequency $s$.  In other words, the
resulting solution for the response ${\bf a}$ given the perturbation
${\bf b}$ takes the form ${\bf a} = {\cal R} (s) {\bf b}$ for We may
straightforwardly include the self-gravity in the response by noting
the a self-gravitating response is the simultaneous solution of the
system to both the perturbation and the response of the system to its
own response.  Mathematically, this is ${\bf a} = {\cal R} (s) \left(
  {\bf a} + {\bf b} \right)$ which upon solving for ${\bf a}$ yields
${\bf a} = \left[{\bf 1} - {\cal R} (s)\right]^{-1} {\cal R} (s) {\bf
  b}$

Note our accomplishment: we began with a coupled set of
partial-integrodifferential equations and end up with a matrix
inversion.  The computational work is all in determining the matrix
elements of ${\cal R}$.  Finally, after solving these sets of linear
equations, we perform an inverse transform to obtain the resulting
response to the perturbation in physical space.

I feel that the name {\em matrix method}\index{matrix method} is a bit
of a misnomer, or at least not fully descriptive.  The procedure
described above has simple intuitive interpretation and this be even
more apparent as we proceed through the next example.  In transforming
to Fourier space, we are in essence solving for the spectrum of normal
modes\index{normal modes} of the system.  The perturbation picks out
the discrete modes and excites ``packets'' of continuous modes.  After
transforming back to physical space, we see the result of the decaying
(or growing) discrete modes and phase-mixing packets of continuous
modes in configuration space.  In this sense, this approach might be
more aptly called {\em stellar spectral dynamics}\index{stellar
  spectral dynamics}.

\subsection{Example: slab dispersion relation}
\label{sec:slabdisp}

In this section, we apply this {\em spectral dynamics} approach to
stellar slab described in \S\ref{sec:slabex}.  The natural coordinates
here are Cartesian.  The canonical variables describing the phase
space are linear position and momentum in the slab and action-angle
variables in the vertical direction.  This simple case differs from a
disk or halo in that trajectories are not bound in the two in-plane
dimensions.  Similarly, there is symmetry in the two in-plane
dimensions so with no loss of generality we are free to consider only
one of these, say the $x$ degree of freedom. So, the canonical
variables are linear momentum and position ($p_x, x$) and vertical
action and orbital angle ($I_z, \theta$).  Orbital angle is defined
as:
\[
\theta = \Omega_z \int^t_0 dt = \Omega_z \int^z_0 {dz \over
  \sqrt{2(E_z-\Phi(z))}}
\]
where $E_z$ is the energy in the vertical degree of freedom and
$\Omega_z(E_z) = {\partial H/\partial I_z}$.  The density and
potential of the unperturbed equilibrium model does not vary in the
infinite horizontal plane so the unperturbed quantities---density,
potential and phase-space distribution function---do not vary in this
dimension.  This presents a formal difficultly popularized by Binney
\& Tremaine as the ``Jeans' swindle''.  We will side step the
subtleties here but please see Binney \& Tremaine
(1987\nocite{Binney.Tremaine:87}) for discussion. We can now write our
linearized equations of motion, the CBE and Poisson equation in these
variables:
\begin{equation}
  {\partial f_1\over\partial t} + {\partial
    f_1\over\partial\theta}\Omega_z(E_z) + {\partial f_o\over\partial
    x}p_x - {\partial f_o\over\partial E}\,{\partial
    V_1\over\partial\theta}\Omega_z(E_z) - {\partial f_o\over\partial
    p_x}{\partial V_1\over\partial x} = 0, \quad
  \nabla^2V_1({\bf r}) = 4\pi G\rho_1({\bf r}).
\end{equation}
We now perform the two transforms: Fourier in actions and angles and
Laplace in time.  Again, the infinite horizontal extent causes a
slight complication: a continuous set of plane waves rather than a
discrete set that would obtain from a bound system.  Let us denote the
Fourier wave vector in the $x$ direction as ${\bf k}$, the index of
the discrete vertical set as $n$ and the Laplace variable as $s$.  A
tilde indicates a Laplace transformed quantity.  The transformed CBE
becomes
\begin{equation}
  s{\tilde f}_{1\,n{\bf k}} + in\Omega_z{\tilde f}_{1\,n{\bf k}} +
  i{\bf k}\cdot{\bf p}{\tilde f}_{1\,n{\bf k}} - {\partial
    f_o\over\partial E_z}\,i n {\tilde V}_{1\,n{\bf k}}\Omega_z -
  {\partial f_o\over\partial E_x}\,i{\bf k}\cdot{\bf p} {\tilde
    V}_{1\,n{\bf k}} = 0.
\end{equation}
Solving for ${\tilde f}$, we now integrate over velocities to derive
the Laplace-transformed density for each wave vector and vertical
index.  Integrating over wave vectors and summing over vertical
indices gives us the expression for the response density for each
Laplace frequency $s$:
\begin{equation}
  {\tilde\rho}({\bf r}, s) = \sum_{n^\prime} \int d^3k^\prime \int d^3v\,
  {\tilde f}_{n^\prime\,{\bf k}^\prime}
  e^{in^\prime\theta(z)} e^{i{\bf k}^\prime\cdot{\bf R}}.
  \label{eq:tilderho}
\end{equation}
Next, we incorporate the Poisson equation by expanding the density and
potential distortions in the biorthogonal functions in canonical variables::
\begin{equation}
  p_{\mu\,{\bf k}} = {1\over 2\pi}\ 
  e^{i{\bf k}\cdot{\bf R}} \sum_n w_{\mu\,n} e^{in\theta}
  \hbox{\rm\ \ where\ \ }
  w_{\mu\,n} \equiv {1 \over 2\pi }
  \int_0^{2\pi} d\theta\, \Psi_\mu(z)e^{-in\theta}.
\end{equation}
Using the biorthgonality condition we perform the scalar product with
equation (\ref{eq:tilderho}) to get an linear set of equations that
determine the expansion coefficients
\begin{equation}
  a_{\,\mu{\bf k}} = -4\pi G \int d^3x\,d^3v\,  
  p_{\mu\,{\bf k}}^\ast
  {\tilde f}_{n\,{\bf k}}
  e^{in^\prime z} e^{i{\bf k}^\prime\cdot{\bf R}}.
\end{equation}
Substituting the solution for ${\tilde f}$, we have explicitly
\begin{equation}
  a_{\mu\,{\bf k}}
  = -4\pi G (2\pi) \sum_{n,\nu} \int dv_x\,dv_y \int dI_z \,
  {
    \,{\partial f_o/\partial E_z}\, n \Omega_z + 
    {\bf k}\cdot{\bf p} \,{\partial f_o/\partial E_x}
    \over 
    n^\prime \Omega_z + {\bf k}\cdot{\bf p} - i s
    }
  w_{\mu\,n}
  w_{\nu\,n}
  \, b_{\nu\,{\bf k}}
  \label{eq:matrixeqn1}
\end{equation}
which can be written as the following matrix equation
\begin{equation}
  a_{\mu\,{\bf k}} \equiv  \sum_n \sum_\nu M^n_{\mu\nu}(s) \,
  b_{\nu\,{\bf k}}
  \label{eq:matrixeqn2}
\end{equation}

To get the full self-gravitating response, we note that the imposed
perturbation is then the sum of the internal response and external
perturbation as follows:
\begin{equation}
a_{\mu\,{\bf k}} = \sum_n \sum_\nu M^n_{\mu\nu}(s,{\bf k}) \, \left(
  a_{\mu\,{\bf k}} + b_{\nu\,{\bf k}} \right).
\end{equation}
The solution for the response is then
\begin{equation}
a_{\mu\,{\bf k}} = \left[1 - M^n_{\mu\alpha}\right]^{-1} {\cal
  M}_{\alpha\nu} b_{\nu\,{\bf k}} = {\cal D}_{\mu\alpha}^{-1} {\cal
  M}_{\alpha\nu} b_{\nu\,{\bf k}}.
\end{equation}
Alternatively, we can look for the perturbation that has the same
shape as its own response, an eigenmode.  The equation for this
solution takes the form: $a_{\mu\,{\bf k}} = {\cal M}_{\mu\nu}
a_{\nu\,{\bf k}}$.  A non-trivial solution demands that ${\cal D}(s)
\equiv \det\{{\bf 1} - {\cal M}_{\mu\nu}(s)\}=0$ and this is often
called the {\em dispersion relation}\index{dispersion relation} by
analogy with the same relation that defines the possible wave modes in
a plasma.  We can classify the resulting modes by the real part of
$s$.  If $\Re(s)>0$, $\Re(s)=0$ and $\Re(s)<0$, the mode is growing,
oscillatory and damped, respectively. If we are interested in the
evolution of a stable system, growing modes should be absent from the
spectrum by design.  Oscillatory modes are rare, requiring pattern
frequencies which avoid commensurabilities with an integer
combination of orbital frequencies.  For reason, pure oscillating
modes are practically non-existent, although one can construct special
cases theoretically.  The damped part of the spectrum has analogy with
Landau damping in a plasma.  Physically, the damping results from
resonant transfer between the pattern and commensurabilities with
orbital frequencies.

Note that all of these solutions are in Laplace space.  To recover the
time evolution, we must perform the inverse Laplace transform.  This
requires a bit of care but is straightforward (see the standard plasma
literature, e.g. Krall \& Trivelpiece, 1973 or Ikeuchi \& Nakamura
1974 for details\nocite{Krall.Trivelpiece:73,Ikeuchi.Nakamura.ea:74}).

\def\erf{\mathop{\rm erf}\nolimits} 
\def\Res{\mathop{\rm Res}\nolimits}

Finally, let us evaluate the response explicitly for a specific case.
Recall that we are assuming that Let ${\bf{\hat e}}_k$ is ${\bf{\hat
    e}}_x$.  Let us further assume that we can factor the phase-space
distribution function as: $f_o(z, {\bf v}) = f_\parallel(v_x, v_y)
f_\perp(z, v_z)$ Let the in-plane part of the distribution function be
Maxwellian and the vertical part be be that for the $\sech^2(z/h)$
density profile.  The matrix elements $M^n_{\mu\nu}(s,{\bf k})$ now
take the form
\begin{equation}
M^n_{\mu\nu}(s,{\bf k}) = -4\pi G (2\pi) \int dI_z \,
\int dv_x
{
  \,{\partial f_\perp/\partial E_z}\, n \Omega_z -
  kv_x/\sigma^2 f_\perp
  \over 
  n^\prime \Omega_z + {\bf k}\cdot{\bf p} - i s
  } 
{ e^{-v_x^2/2\sigma^2} \over \sqrt{2\pi\sigma^2} }
w_{\mu\,n}
w_{\nu\,n}
\end{equation}

Conveniently, the integrals over $v_x$ can be written as error
functions of complex argument using the relation
\begin{equation}
  \int^\infty_{-\infty} dv {e^{-v^2/2\sigma^2} \over kv + q} = 
{1\over k}
\int^\infty_{-\infty} dy {e^{-y^2} \over y + z} = -\pi i \erf(-iz) e^{-z^2}
\end{equation}
where $z=(n\Omega_z - is)/k\sqrt{2\sigma^2}$.  Routines for evaluating
the complex error function are readily available (e.g. {\tt
  www.netlib.org}).  Now let's look at a few applications.

\subsection{Modes in the slab}

\begin{figure}
  \mbox{
    \includegraphics[width=2.6in]{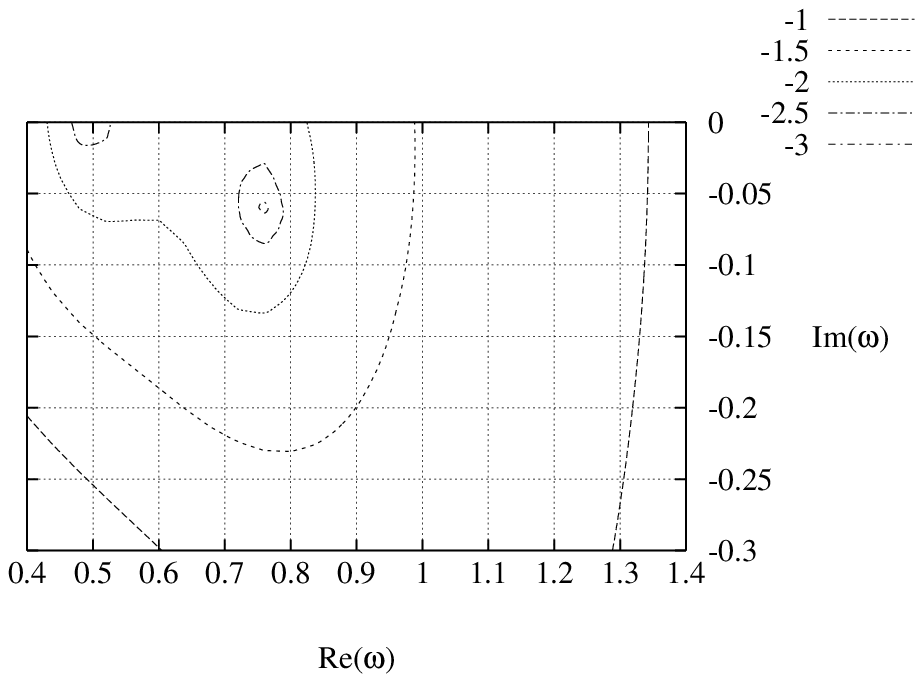}
    \includegraphics[width=1.8in]{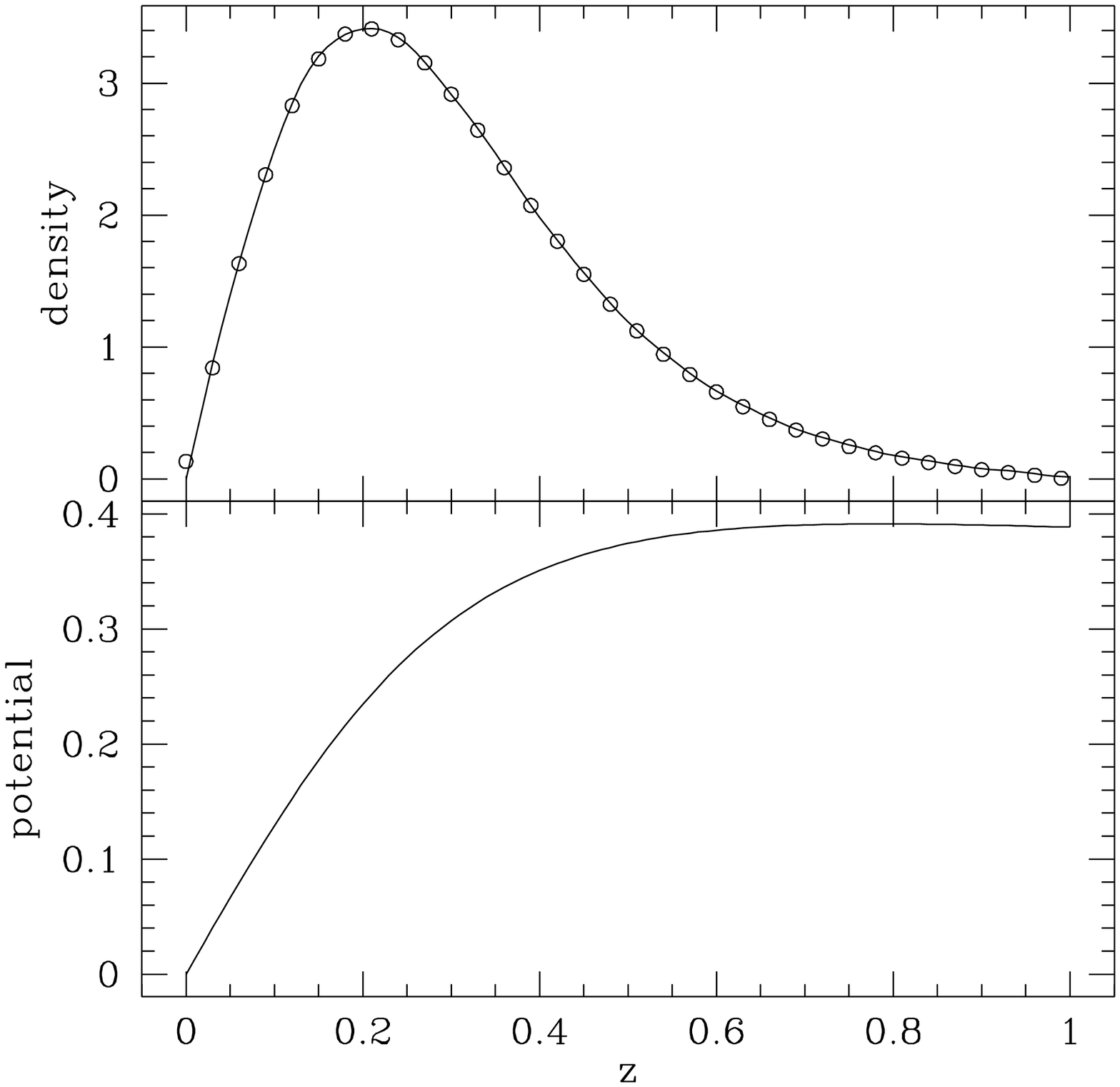}
    \includegraphics[width=1.8in]{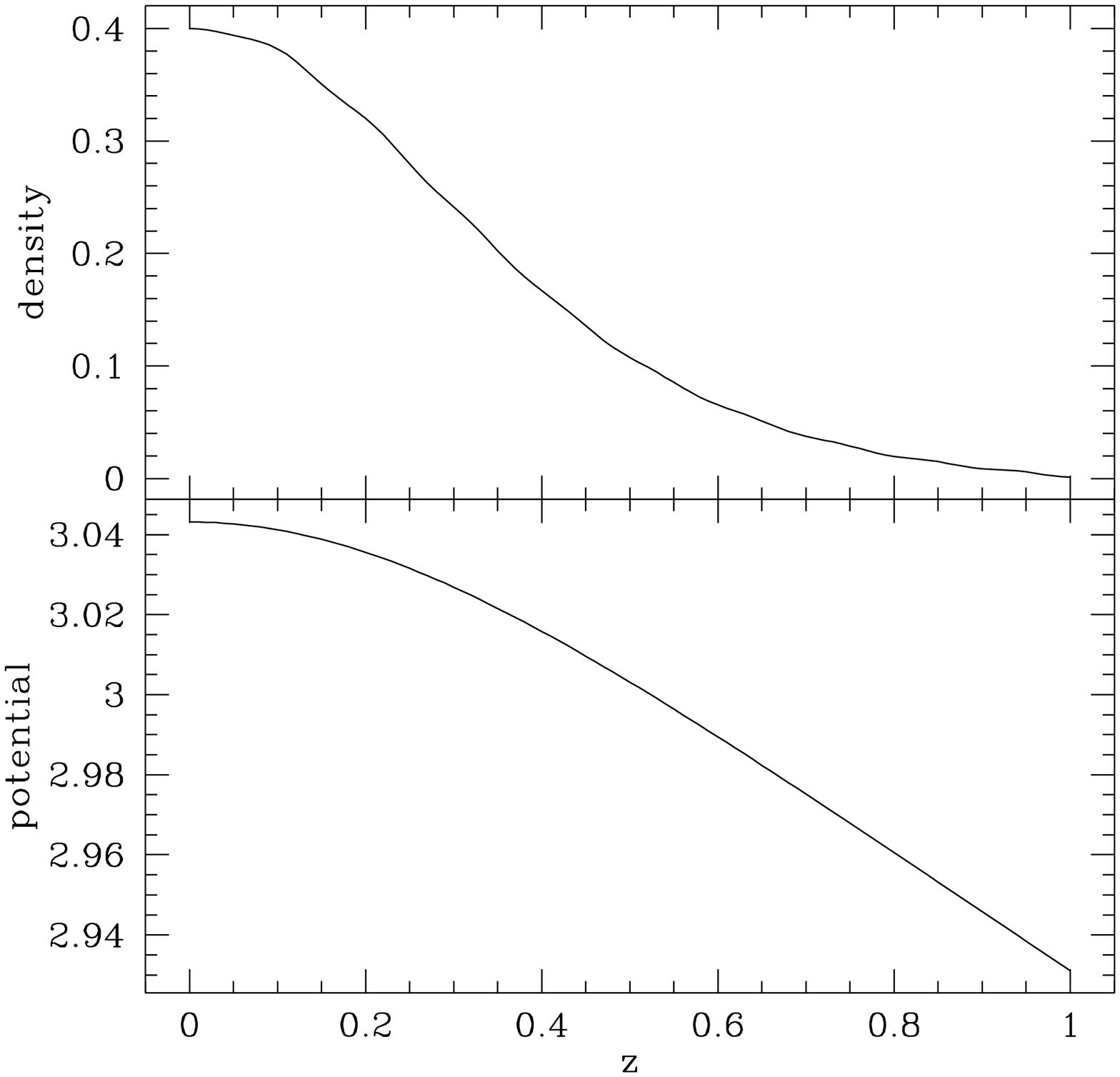}
    }
  \caption{Left: Plot of the dispersion relation $|D(\omega)|$ 
    for the slab with a Maxwellian and sech-squared distribution.
    Center: odd mode (bending) and right: Even mode (Jeans).}
\label{fig:slabfreqs}
\end{figure}

Modes are at the zeros of ${\cal D}$ which is shown in Figure
\ref{fig:slabfreqs}.  The figure only shows the dispersion relation as
a function of $\omega\equiv-is$ rather than $s$. For reference,
$\Im(\omega)>0$ corresponds to instability.  The dispersion relation
is even in $\Re(\omega)$ an exploration of the half-plane
$\Re(\omega)>0$ is sufficient.  We see two zeros.  The first has
$\Im(\omega)<0$ and very small $|\Im(\omega)|\ll 1$.  This is a damped
mode\index{weakly damped mode} but very weakly damped.  The second,
with larger $\Re(\omega)$ is also weakly damped more strongly that the
first.

To get physical intuition for these modes, one can can determine the
shape of the mode by finding the null vector of $M^n_{\mu\nu}(s,{\bf
  k})$ for each zero in Figure \ref{fig:slabfreqs}.  The two modes are
shown also in Figure \ref{fig:slabfreqs}.  The most weakly damped of
the two is odd about the mid plane and is a traveling bending mode.
The second mode mode is even about the mid plane and is a breathing
mode.  The dispersion relation ${\cal D}$ is also a function of $k$.
The zeros ${\cal D}$ determine a branch for each mode.  In this case,
the damping increases as $|k|$ increases.

\subsection{Excitation of a damped mode by a disturbance}

We can use the information about the various modes in the dispersion
to compute the excitation of the system due to a time-dependent
disturbance.  For example, let us consider the response of the slab
due to a body passing through the slab at constant velocity.  This is
an idealization of a dwarf satellite\index{dwarf galaxy} moving
through the disk (e.g. Sgr dwarf and the Milky Way).

We assume that we known time dependent of disturbance to start.  After
expanding this in our chosen biorthogonal basis, we can write this as
vector of time-dependent of coefficients.  The Laplace transform of
the perturbation vector is then
\begin{equation}
  {\bf b}(s) = \int^\infty_0 dt^\prime
  \exp(-st^\prime) {\bf b}(t^\prime)
\end{equation}

The inverse Laplace transform of equation (\ref{eq:matrixeqn1}) gives
\begin{eqnarray}
  {\bf a}(t) &=& {1\over2\pi i} \int^{c+i\infty}_{c-i\infty} ds\,e^{st}
  {\cal D}_{\mu\alpha}^{-1}(s,{\bf k}) {\cal
    M}_{\alpha\nu}(s,{\bf k}) \int^\infty_0 dt^\prime
  \exp(-st^\prime) {\bf b}(t) \nonumber \\
  &=& \int^\infty_0 dt^\prime
  {\bf b}(t^\prime) {1\over2\pi i} \int^{c+i\infty}_{c-i\infty}
  ds\,e^{s(t-t^\prime)} {\cal D}_{\mu\alpha}^{-1}(s,{\bf k}) {\cal
    M}_{\alpha\nu}(s,{\bf k})
\end{eqnarray}

The Laplace transform was performed assuming a value of $s$ that
insured convergence.  We are free to deform the integral path as long
as we use care to analytically continue the integrand and identify
singularities.  In particular, if the slab is dynamically stable, then
${\cal D}$ is non-singular in the half plane with $\Re(s)>0$.  There
will be poles for $\Re(s)<0$ corresponding to damped modes.  In
addition, the matrix elements ${\cal M}$ have denominators of the form
$s+ix$ for $x$ on the real line.  The contour deformation rules are
then: (1) for $t<t^\prime$, deform to $\Re(s)\rightarrow\infty$, no
poles; and (2) for $t>t^\prime$, deform to $\Re(s)\rightarrow-\infty$,
poles at $s=-ix$ and at any possible poles of ${\cal D}$ in the
lower-half $s$-plane (damped modes).  Performing the inverse Laplace
transform and putting everything together gives the explicit
expression for the self-gravitating time-dependent response to the
perturbation:
\begin{eqnarray}
  {\bf a}(t) &=&
  -4\pi G (2\pi) \int dI_z \,\int dv_x\,i\left(
    {\partial f_o\over\partial E_z}\, n \Omega_z + kv_x {\partial f_o\over\partial
      E_x} \right) \times \nonumber \\ 
  && {\cal D}_{\mu\alpha}^{-1}(-ix,{\bf k}) w_{\mu\,n} w_{\nu\,n}
  \int^\infty_0 dt^\prime
  {\bf b}(t^\prime) e^{-ix(t-t^\prime)} + \nonumber \\
  && \sum_{s_r} \Res {\cal D}_{\mu\alpha}^{-1}(s_r,{\bf k})
  {\cal M}_{\alpha\nu}(s_r,{\bf k}) \int^\infty_0 dt^\prime
  {\bf b}(t^\prime) e^{s_r(t-t^\prime)}
  \label{eq:slabresp}
\end{eqnarray}
The inverse of the dispersion matrix will have poles at any modes
(recall Cramer's formula).  The notation $\Res {\cal
  D}_{\mu\alpha}^{-1}$ denotes the residue of the this matrix and may
be determined numerical using singular value decomposition with the
following procedure: (1) locate the damped modes $s_r$ and compute
${\cal D}_{\mu\nu}(s_r)$; (2) analyze by singular value decomposition
and compute the determinant without the singular value, $D\prime(s_r)$
say; (3) compute the derivative of the determinant at $s_r$,
$dD/ds|_{s_r}$. We expect $D(s) = a(s-s_r) D^\prime(s_r)$ for some
unknown constant of proportionality whose solution is:
$a=dD/ds|_{s_r}/D^\prime(s_r)$; and (4) replace the singular value in
the decomposition by the value of $a$.  I have given explicit details
for readers interested in exploring this procedure numerically. The
numerical computations here are straightforward for this case of the
slab.  One should be able to investigate the full response of the slab
to an arbitrary perturbation.

\section{Galaxy interactions}

Let us finish with examples of these methods applied to two classes
astronomical scenarios.  First, we will mention the excitation of
structure by a passing galaxy such as a weak encounter in group, a
{\em fly-by}.  These interactions can cause off-centered disks and
centers and trigger bars.  Similarly, an orbiting satellite will have
a very similar effect on its primary.  Second, we will describe
noise-driven evolution, both the shape and magnitude of
fluctuation-driven structure and the possibility of significant
evolution of halo profiles due to these fluctuations.

\subsection{Fly-bys and satellites}

Another way of getting the same sort of excitation, perhaps more
important for group galaxies than the Milky Way, is a passing
fly-by\index{fly-by encounter}.  A perturber on a parabolic or
hyperbolic trajectory can excite similar sorts of halo asymmetries and
persist until long after the perturber's existence is unremarkable.
Presumably, our Galaxy has suffered such events in the past but
because the satellite excitation is closely related to the fly-by
excitation, the study of one will provide insight into the other.
Vesperini \& Weinberg \cite{Vesperini.Weinberg:00} describes the
application of the response approach to this problem.  From these
analytic calculations, we can compute the standard asymmetry
parameters \cite{Abraham.etal:96a,Abraham.etal:96b,Conselice.etal:00}
obtained by summing over the mean square difference of the galaxy and
its $180^\circ$ rotated image:
\begin{equation}
  A = {1\over2} {\sum\left|I(x,y) - I_{rot}(x,y)\right|\over \sum I(x,y)}.
  \label{eq:Adef}
\end{equation}
For example, a perturber with 10\% of the halo mass, with pericenter
at the halo half-mass radius, and encounter velocity of 200 km/s will
produce $A\approx 0.2$.  Damped modes play a major role in both the
morphology and longevity of these modes.  Figures 6 and 7 in Vesperini
\& Weinberg (2000) illustrates their importance by comparing the
response with and without damped modes.  The $m=1$ mode is
significantly altered by the discrete weakly damped mode.  Please see
\cite{Vesperini.Weinberg:00} for more details.

Because the halo response is dominated by the modes of the halo rather
than properties of the perturber, we expect that the asymmetry should
be dominated by contributions at well-defined radii, independent of
the perturber parameters.  We proposed a simple generalization of
equation (\ref{eq:Adef}) to test this prediction: define $A(r)$ to be
the sums over pixels restricted to those within projected radius $r$.
More recently, we have shown that n-body simulations agree in
magnitude and morphology with the perturbation theory.

\subsection{Noise}
\label{sec:noise}

The possibility of long-lived damped modes leads to the possibility
that global modes are continuously excited by a wide variety of events
such as disrupting dwarfs on decaying orbits, infall of massive high
velocity clouds, disk instability and swing amplification and the
continuing equilibration of the outer galaxy.  The dominant halo modes
are low frequency and low harmonic order and therefore can be driven
by a wide variety of transient noise sources.  Some recent work
\cite{Weinberg:00a,Weinberg:00b} provides a theory excitation by noise
and applies this to the evolution of halos.  In this section, I will
first describe an application of our response theory to fluctuation
noise.  I will describe some preliminary results suggest that noise
may drive a halos toward approximately self-similar profiles.
Additional work will be required to make precise predictions for these
trends and explore the consequences for long-term evolution of disks
in spiral systems.

\subsection{Halo noise}

\begin{figure}
  \mbox{
    \includegraphics[width=3.0in]{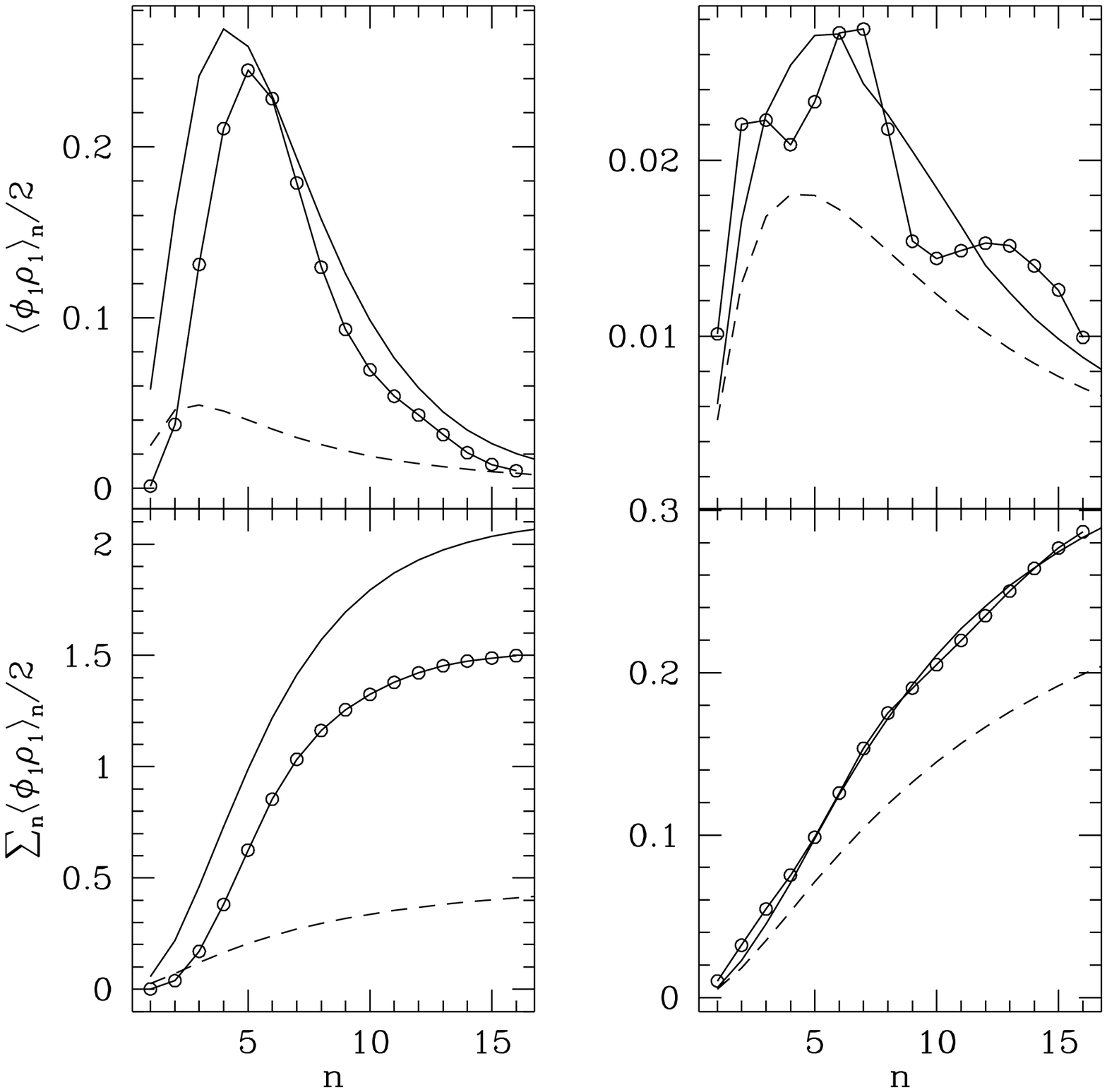}
    \includegraphics[width=3.0in]{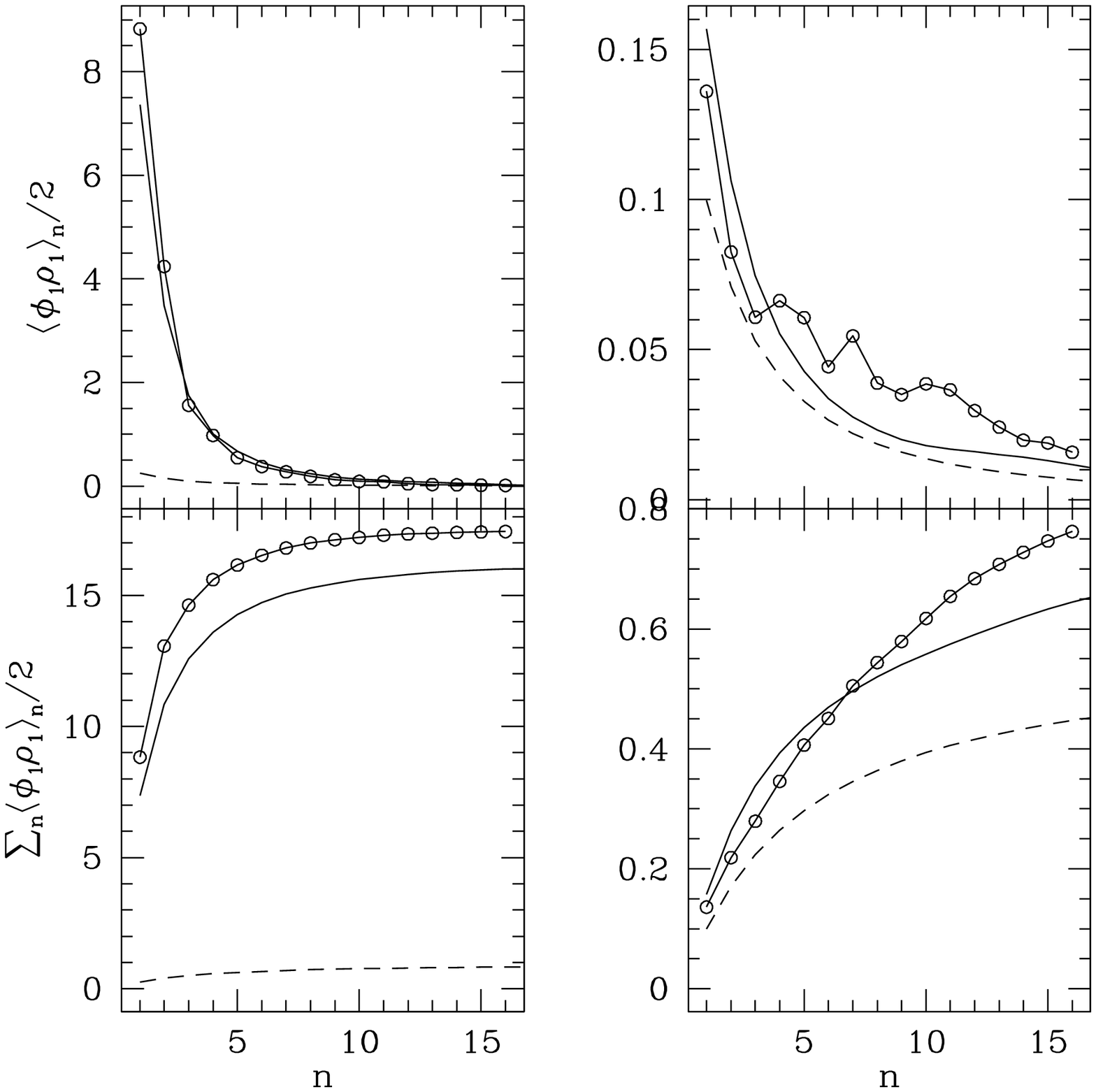}
    }
  \caption{Power (in energy units) of the response of a halo to noise
    for two different models as a function of radial basis index.
    Left: $W_0=5$ King model. Right: Hernquist model.  The top row
    left (right) shows the $l=1$ ($l=2$) response for each model.  The
    bottom row shows the cumulative power.  The radial basis set is
    similar to that shown in Fig. \protect{\ref{fig:converge}}; the
    index on the abscissa indicates the number of nodes for each basis
    function.  The larger the index, the finer the spatial scale.}
  \label{fig:fluctmodels}
\end{figure}

The simplest approach is a calculation of the power in a stellar
system due to Poisson fluctuations\index{halo noise}.  Consider the
response of the entire system to a single orbiting star.  Physically,
each star excites a wake in the halo.  This wake includes all the
modes from the weakly damped modes to very small scale modes.  We now
sum up the wakes from all of the stars.  The self-gravity of the
lowest-order mode leads to significant excess power at large scales.
The detailed theoretical computation is compared with n-body
simulations in Figure \ref{fig:fluctmodels}.  Note that the
amplification of the noise by self-gravity is significant for the
$l=1$ component for both halos with and without cores.
  
\begin{figure}
  \mbox{
    \includegraphics[width=3.0in]{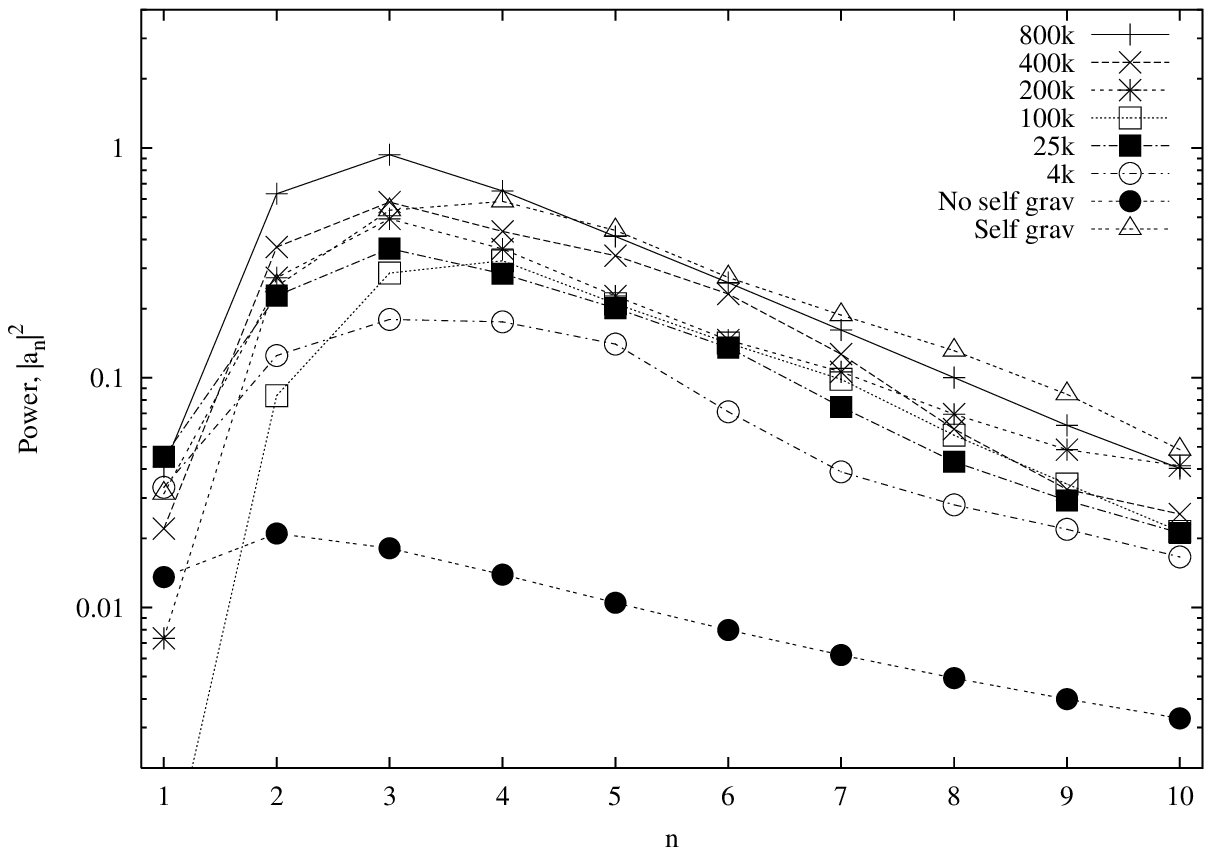}
    \includegraphics[width=3.0in]{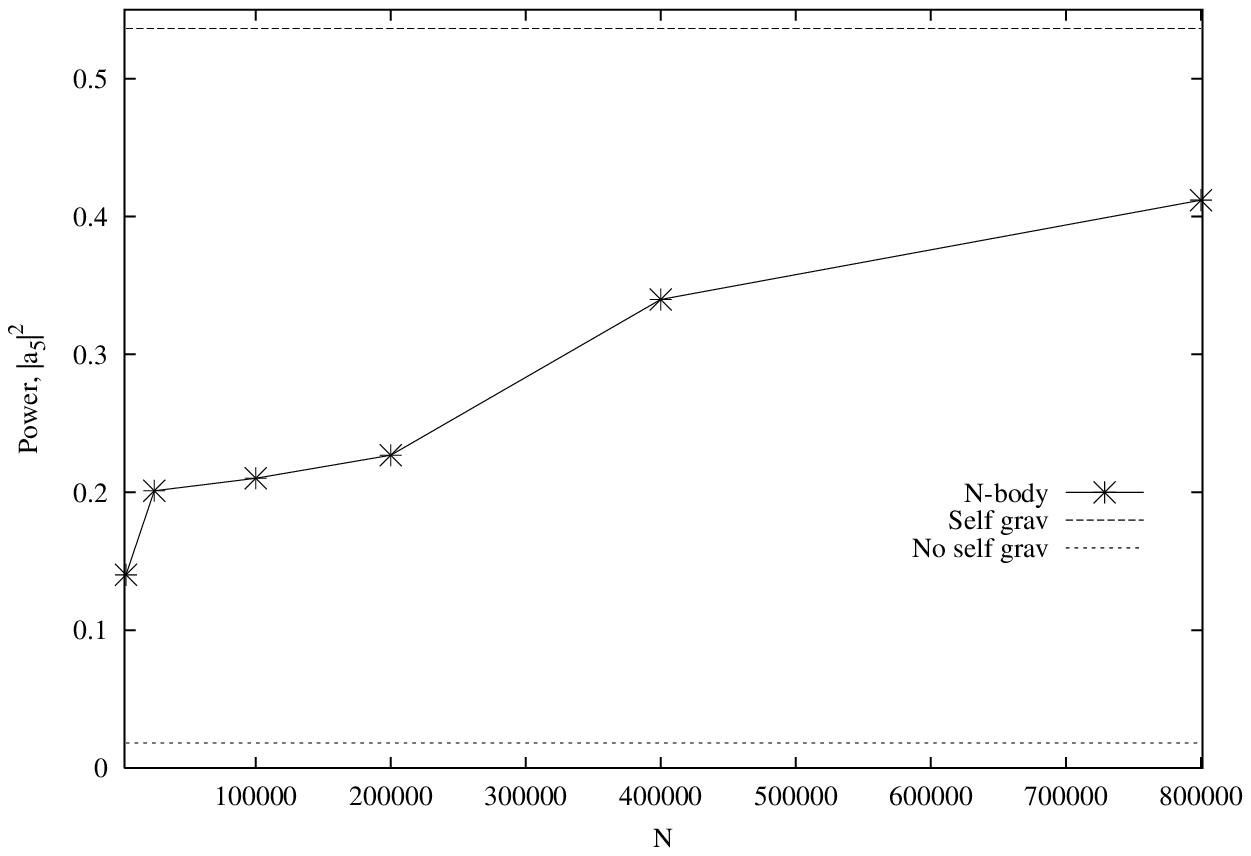}
    }
  \caption{Left: Fluctuation power as a function of particle number
    for each basis coefficient scaled by the number of particles $N$.
    Right: Fluctuation power for $n=5$ as a function of particle
    number.  The upper (lower) horizontal lines show the expected
    results with self gravity (Poisson).  }
  \label{fig:fluctN}
\end{figure}

The analytic calculation is valid in the limit $N\rightarrow\infty$.
However, if this is not obtained in the n-body simulation, the power
in fluctuations can be so large that individual orbits do not have
well-defined conserved quantities (energies and angular momenta) over
a dynamical time.  In this {\em noise-dominated} regime, the diffusion
of orbits is so fast that coherent large-scale dynamics is suppressed.
In other words, with too few particles, one is simulating a star
cluster not dark-matter dominated halo.  We can see the effects of
particle number by determining the number $N$ required to obtain the
noise-spectrum predicted by analytic solution of the underlying power
spectrum.  This is illustrated in Figure \ref{fig:fluctN}.  The figure
compares same empirically determined power spectra shown in Figure
\ref{fig:fluctmodels} (left panel) but for various values of $N$.  In
short, one needs $N\ge10^6$ before the dynamics of the collisionless
limit obtains.  This result is largely independent of the n-body
simulation technique.

\def\btheta{{\bf\theta}}
\def\bx{{\bf x}}
\def\bI{{\bf I}}
\def\bbI{{\bf{\bar I}}}
\def\bxi{{\bf\xi}}
\def\bOmega{{\Omega}}           

\subsection{Evolution of galaxy by noise}

Given that fluctuations are a generic part of stellar dynamics, let us
know ask what sort of evolution we can expect.  To do this, I will
sketch the development of a constitutive equation for the long-term
evolution under noise.  We could proceed as for globular clusters:
expand the Boltzmann collision term using Master formalism
\cite{Binney.Tremaine:87}.  After a number of false starts, I found
the more general transition probability approach to be more natural
(although the Master approach is formally equivalent\index{Master
  equation}).  One begins with the probability that an orbit with
phase-space state $\bx$ at time $t$ makes a transition to $\bx^\prime$
at time $t+\tau$: $P(\bx^\prime, t+\tau|\bx, t)$.  For the entire
ensemble described by the distribution function $f(\bx, t)$, we can
describe the evolution using the transition probability as
\begin{equation}
  f(\bx, t+\tau) = \int d\bx^\prime  P(\bx, t+\tau|\bx^\prime, t)
  f(\bx^\prime, t). 
\end{equation}
Now, expand the transition probability in its moments of $\bx-\bx^\prime$
for small $\tau$.  This gives
\begin{equation}
  {\partial f(\bx, t)\over\partial t} = 
  \sum_{n=1}^\infty \left(-{\partial\over\partial\bx}\right)^n
  D^{(n)}(\bx, t) f(\bx, t)
\end{equation}
This is known as the {\em Kramers-Moyal expansion}\index{Kramers-Moyal
  expansion}\cite{Risken:89}.

We will derive the transition probability for our case by considering
the change in conserved quantities of orbits (actions) over the
correlation time of the fluctuation.  This implies that the transition
probability is only defined for time scales $\tau$ larger than the
dynamical orbital time scales.  Therefore, we can further simplify the
computation by using action-angle variables and averaging over the
rapidly varying angles.  For the phase-space distribution function,
the Kramers-Moyal expansion becomes
\begin{equation}
  f(\bI, t+\tau) = \int d\bI^\prime\,
  P(\bI, t+\tau | \bI^\prime, t) f(\bI^\prime, t)
\end{equation}
Now to evaluate this equation, expand integrand in a Taylor series
about $\bI$ and define $\Delta\equiv\bI^\prime - \bI$.  In the limit
$\tau\rightarrow0$, we
\begin{equation}
  {\partial f(\bI, t+\tau)\over\partial t} = 
  \sum_{n=1}^\infty \left(-{\partial\over\partial \bI}\right)^n
  D^{(n)}(\bI, t)   f(\bI, t) .
\end{equation}
where $D^n$ is proportional to the time-derivative of the moments of
$\Delta$ over the distribution $P$.
However, despite the appearance of continuous functions in these
formulae, note that $P$ describes stochastic events.  To write this
explicitly in stochastic variables, let $\bxi$ be the
stochastic value of $\bI$.  The expression $D^{(n)}$ may be written
\begin{equation}
  D^{(n)}(x, t) = \left.{1\over n!}\lim_{\tau\rightarrow0}{1\over\tau}
    \langle[\bxi(t+\tau) - \bI]^n\rangle\right|_{\bxi(t)=\bI}.
\end{equation}

If stochastic excitation is a Markov process, this guarantees that
the expansion terminates after two terms \cite{Pawula:67}.  Our
evolution equation is then a Fokker-Planck equation:
\begin{equation}
  {\partial f(\bI, t)
    \over
    \partial t} =
  \left\{
    -{\partial\over\partial\bI_i} D^{(1)}_i(\bI, t) + 
    {\partial^2\over\partial\bI_i\partial\bI_j} D^{(2)}_{ij}(\bI, t)
  \right\}
f(\bI, t).
\label{eq:fp}
\end{equation}

\subsection{Noise-dominated halo evolution}

To end this section, we will describe the application of this
formalism to the long-term evolution of a halo under a several
representative noise processes.

First, some general observations.  Noise from periodically orbiting
bodies do not give rise to long-term evolution, even though they do
give rise to significant orbital diffusion (as described above).  This
is easily argued.  Changes over long time periods, so-called {\em
  secular} changes, will only occur if the disturbance presents a
torque.  Consider the mean density of an orbiting body over many
dynamical times, for example.  It will only present a torque if it is
a closed, resonant orbit.  At order $l=1$, this requires that the
radial and azimuthal frequencies be equal, as in a Keplerian orbit.
For most halo profiles, these orbits populate the outer edge and
therefore have little effect.  Similarly, at order $l=2$, we add the
possibility of closed, stationary bar-like orbits that have radial
frequencies that are twice the azimuthal frequencies.  This can occur
in homogeneous cores, but these conditions are thought to be rare or
non-existent in realistic halos.  Order $l=3$ is the lowest order that
admits resonant orbits over a wide-range of energies.  This is not
inconsistent with the the results of \S\ref{sec:noise}.  Noise at
orders $l=1,2$ caused by orbiting bodies can cause significant orbital
diffusion without changing the equilibrium profile.  This turns out to
be a corollary of a more general proof of the stability of stellar
equilibria against phase-mixing \cite{Hjorth:94}.  Parenthetically,
N-body folks have used the maintenance of an equilibrium as an
indicator of the collisionless regime.  However, the argument above
shows that the equilibrium will persist even if the rate of orbital
diffusion is high.

Conversely, any transience in the noise source---orbital decay,
fly-bys, disrupting or shearing stellar streams---can excite the
weakly damped modes at low order. Since a galactic halo will suffer
all of these disturbances over its lifetime, direct numerical
estimates suggest that excitation of transient noise will dominate
orbital noise in driving evolution for realistic astronomical
scenarios and I will give examples of these below
\cite{Weinberg:00a,Weinberg:00b}.

The overall procedure is as follows.  We begin with an equilibrium
halo and phase-space distribution function.  To simplify solution of
the Fokker-Planck equation, the distribution is isotropized.  The
evolution equation (\ref{eq:fp}) is now solved in two steps.  First,
we solve the Fokker-Planck equation holding the underlying
gravitational potential fixed for some $\tau$ greater than $1/\Omega$
but small compared to overall evolution time scale.  Second, we ``turn
off'' the collision term and find new self-consistent equilibrium.
The two-step process is repeated to obtain the evolution.

\begin{figure}
\mbox{
  \includegraphics[width=2.1in]{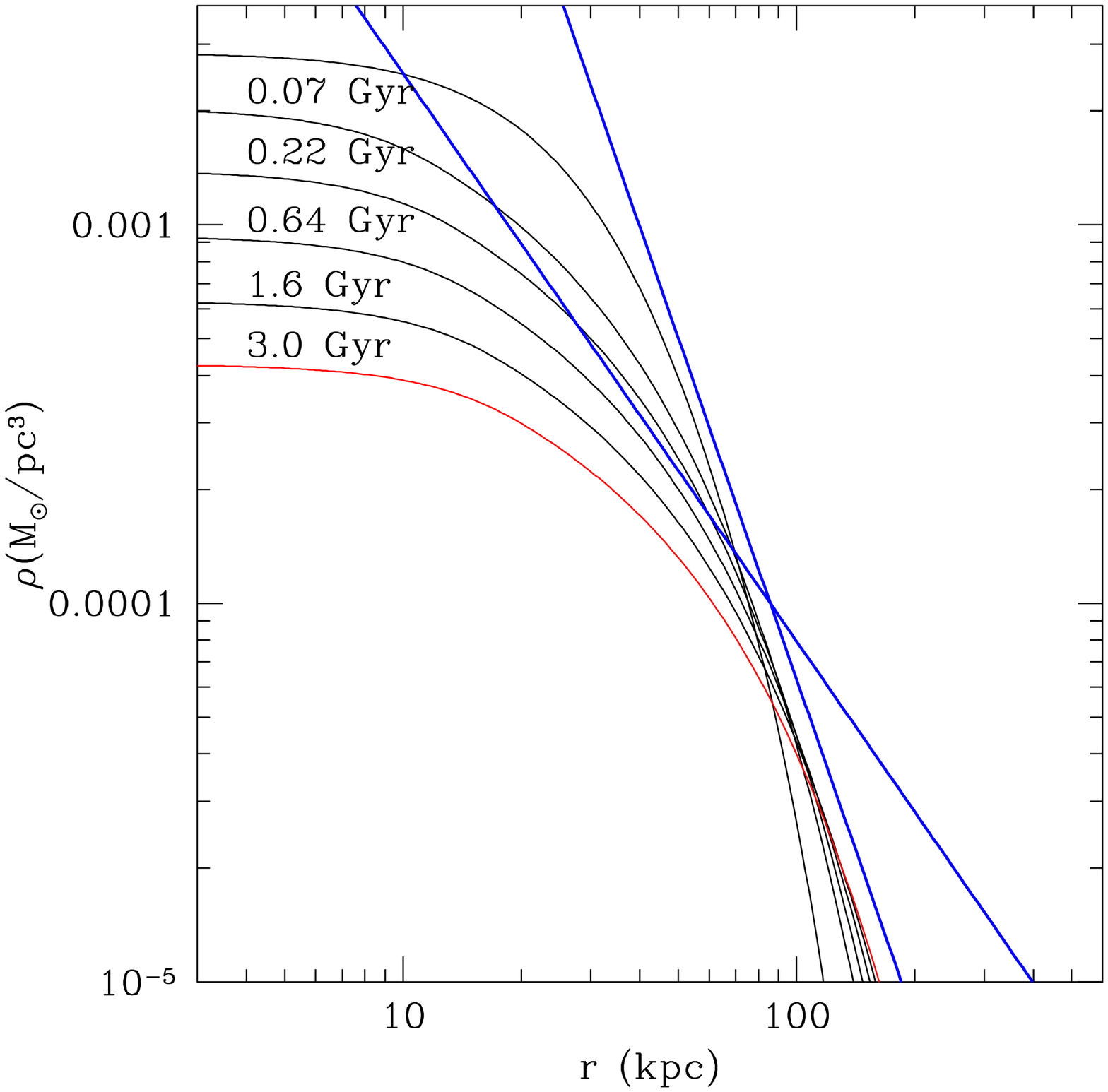}
  \includegraphics[width=2.1in]{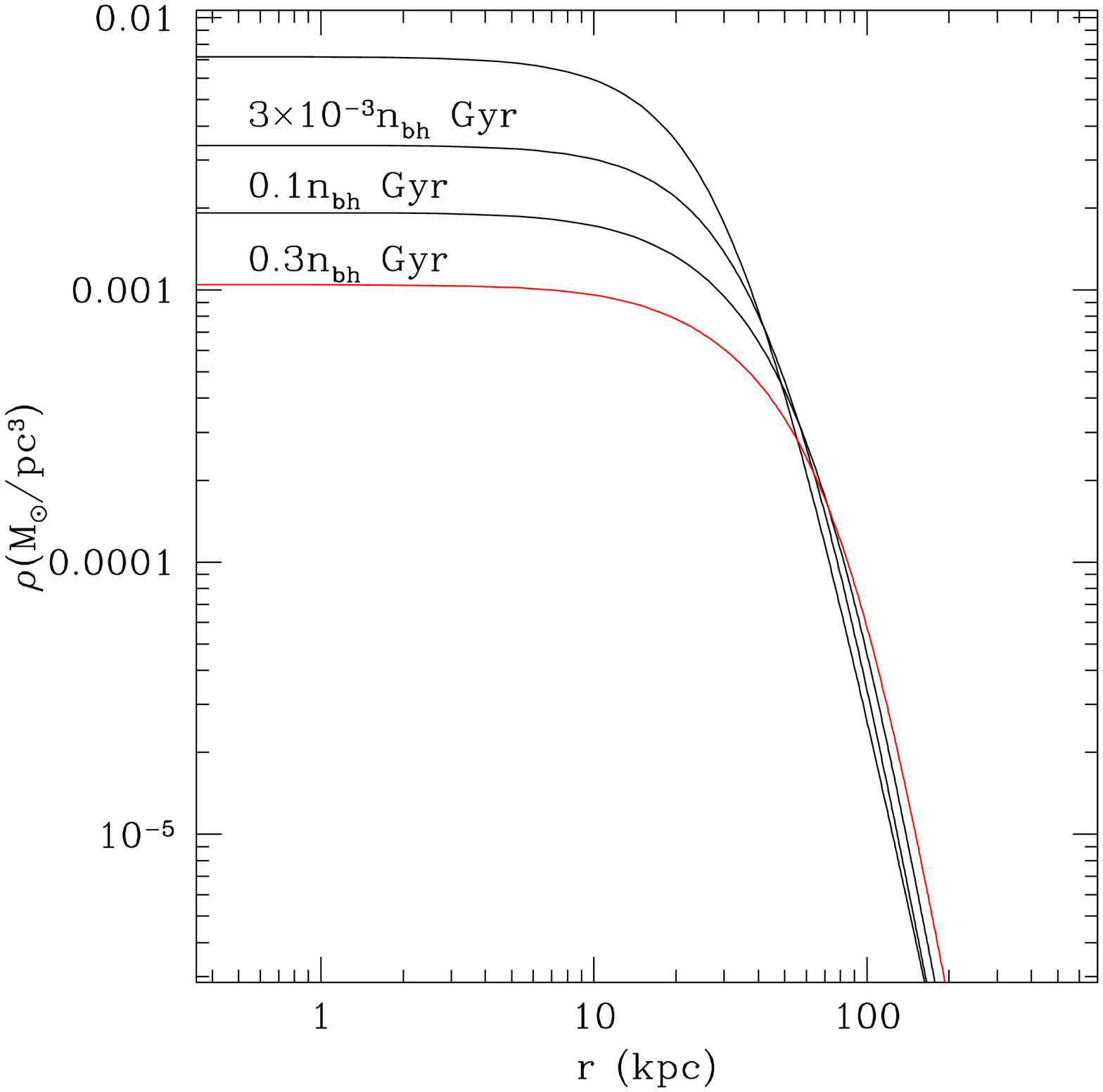}
  \includegraphics[width=2.1in]{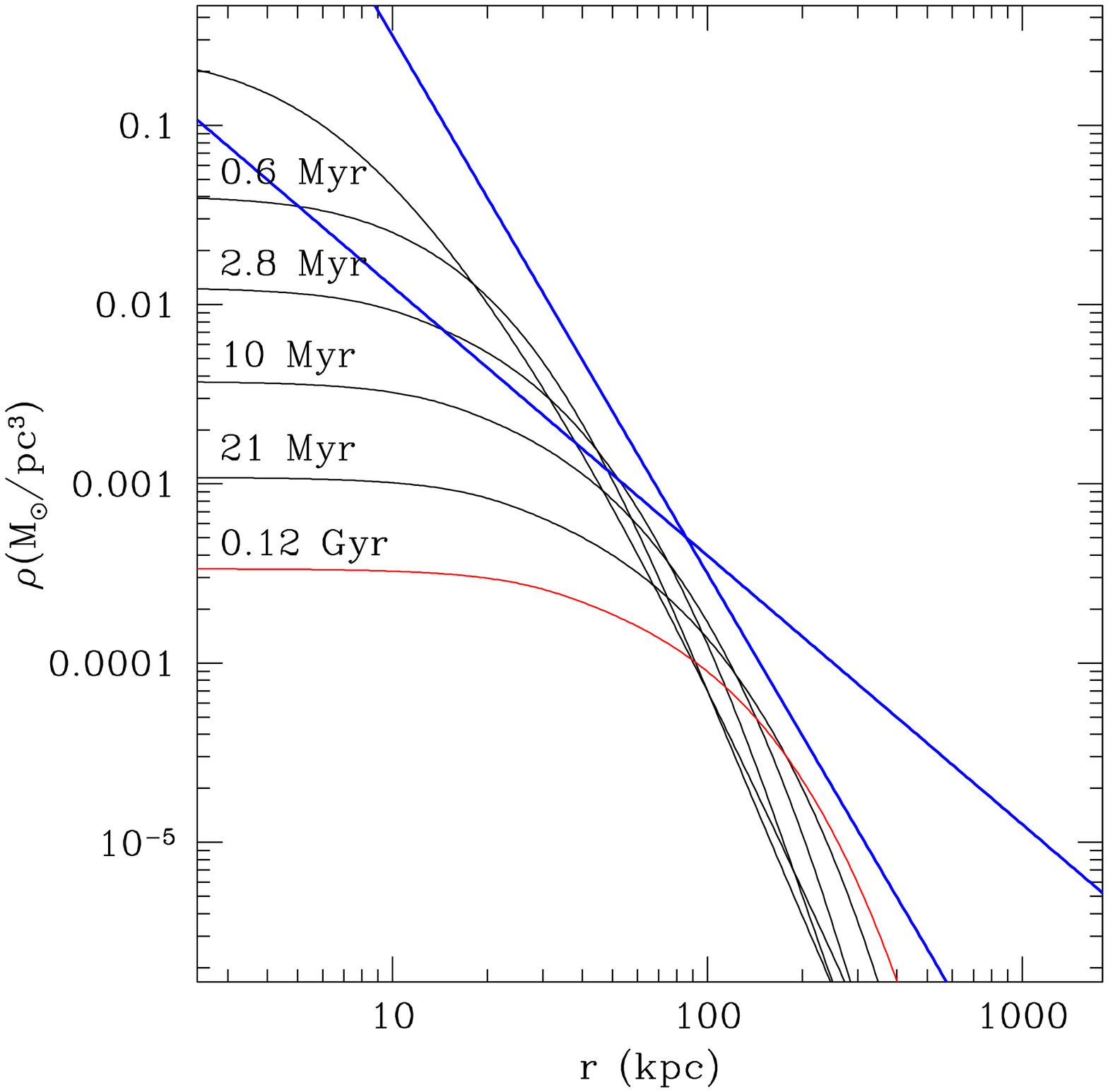}
  }
\caption{Left: Orbital decay in a $W_0=3$ King model halo for a
  satellite to halo mass ratio of 0.05.  The straight lines are power
  laws with exponents $-1.5$ and $-3$, for comparison. Center:
  Evolution of a King $W_0=3$ profile under `black hole' noise.  The
  times for each curve are shown with the scaling for number of black
  holes per halo assuming that the black hole fraction is 10\%.  This
  gives roughly $n_{bh}=10^6$ and the evolution time scale is
  uninterestingly large.  Right: Evolution for a double power law
  model $ \rho \propto (r+\epsilon)^{-\gamma}(r+1)^{\gamma-\beta}$
  with $\gamma=1$, $\beta=4$ and $\epsilon=0.1$.}
\label{fig:haloevolve}
\end{figure}

Figure \ref{fig:haloevolve} shows the evolution under three different
noise sources: (1) a satellite with a decaying orbit; (2) a halo of
black holes; and (3) satellite fly-bys.  In the the first two cases,
we begin with a $W_0=5$ King model and the third begins with a broken
power-law profile (with a small core for numerical convenience).  For
Cases (1) and (3), the results can be characterized as follows.  There
are two distinct evolutionary phases: a transient readjustment to a
double power law profile followed slow, approximately self-similarly
evolution.  The outer profile is characterized by power law with
exponent close to $-3$.  The profile continues to approach the $-3$
power-law form at increasing radius as the evolution continues.
\cite{Weinberg:00b} shows that this obtains for a wide variety of
initial conditions and is caused by the reaction of the halo to the
external $l=1$ multipole, which explains the ubiquity of the profile.
The inner profile has a shallower roll before reaching the core.  A
power law of -1.5 is shown for comparison.  The more concentrated
models, which have deeper potential wells and therefore shorter
dynamical times, evolve most quickly.  This is clear in the comparison
of Cases (1) and (3) but \cite{Weinberg:00b} shows that this obtains
for a variety of initial conditions.  Case (2), evolution by orbiting
black holes, does not result in the same asymptotic form and exhibits
much weaker evolution overall.

Because these models have cores, and both the radial and azimuthal
orbital frequencies are nearly the same in the core, it is difficult
to couple to these orbits in order to transfer angular momentum in and
out of the core.  The core, then, expands with the overall expansion
of the halo due to the deposition of energy from the noise sources.
These dynamics suggest that we restrict our consideration to evolution
beyond the core.  Further investigation of the importance of an
initial cusp are in progress.

\section{Summary and topics for future work}

These lectures have described the use of biorthogonal expansions in
n-body simulations and perturbation theory to understand the long-term
evolution of galaxies.  For a concrete example, I presented an
explicit example of an infinite slab which as a rich modal structure
but can be treated analytically and by n-body simulation with a small
amount of numerical computation.

One can use these same procedures with carefully chosen bases to
represent gravitational field of galaxies to perform smooth,
low-diffusion, n-body simulations.  Multiple disk, bulge and halo
components can be treated simultaneously by using separate bases for
each component since solutions of Poisson equation are additive.  The
same expansion bases can be used to construct perturbation theories
for understanding the stable and unstable modes and deriving the
response to time-dependent disturbances.  The advantage of using the
perturbation theory is its insensitive to particle noise and resulting
orbital diffusion which can wipe out correlations that critical to
dynamics.  Because both the n-body simulations and the perturbation
can be represented by the same field expansion, the two approaches can
be used together to understand the details of a complex interaction.
    
Using these methods, we have seen that many if not all astronomical
equilibria have weakly damped modes.  These modes easy to excite and
slow to decay and therefore will tend to dominate the non-axisymmetric
structure of galaxies.  For example, the ubiquity of very
weakly-damped ``sloshing modes'' ($l=m=1$) may cause lopsided disks,
off-centered nuclei including nuclear bars and black holes.  The basic
dynamics here was throughly explored decades ago by the pioneers in
spiral structure
\cite{Lin.Shu:64,Julian.Toomre:66,Toomre:69,Shu:70a,Shu:70b}.  In
particular, global spiral structure was shown to be damped
\cite{Toomre:69} for the same physical reasons.

We described several applications, satellite and fly-by induced
lopsidedness and bars and excitation of structure by noise,
emphasizing the latter.  In particular, the Poisson noise from a
simulation of a halo with $10^5$ particles drives enough power, when
damped modes are included to cause observable disturbances in the
disk.  Physically, this noise is comparable to a halo of black holes
of 2 to $6\times10^6\msun$.  Conversely, one needs at least $10^7$
bodies to suppress the particle noise to the point that the
collisionless limit is obtained with some confidence.  We then
considered the long-term consequences of this noise to the evolution
of a galaxy halo.  We argued that dwarf mergers, weak encounters with
neighbors, and noise from the still equilibrating outer halo can drive
significant halo evolution through noise excitation over a galaxy
lifetime.

There is much more that needs to be done in this area, including
careful analysis of more realistic galaxy models under a wide variety
of possible perturbations and noise spectra.  Calculations to date
have only considered stellar dynamics, but the gas component response
to the large-scale structure discussed here may prove important to our
understanding of galaxy evolution as well as providing an important
observational diagnostic.  This all leads to the speculative
possibility that galactic evolution may be driven by stochastic
evolution, at least in part.  It will be interesting to see if a
stochastic view rather than static view is borne out.


\begin{susspbibliography}{99}

\bibitem[{Abraham} et~al. 1996a]{Abraham.etal:96b}
{Abraham}, R.~G., {Tanvir}, N.~R., {Santiago}, B.~X., {Ellis}, R.~S.,
  {Glazebrook}, K., and {van den Bergh}, S. 1996a,
\newblock \mnras, 279, L47.

\bibitem[{Abraham} et~al. 1996b]{Abraham.etal:96a}
{Abraham}, R.~G., {van den Bergh}, S., {Glazebrook}, K., {Ellis}, R.~S.,
  {Santiago}, B.~X., {Surma}, P., and {Griffiths}, R.~E. 1996b,
\newblock \apjs, 107, 1.

\bibitem[Arnold 1978]{Arnold:78}
Arnold, V.~I. 1978,
\newblock {\em Mathematical Methods of Classical Mechanics},
\newblock Springer-Verlag, New York.

\bibitem[Binney and Tremaine 1987]{Binney.Tremaine:87}
Binney, J. and Tremaine, S. 1987,
\newblock {\em Galactic Dynamics},
\newblock Princeton University Press, Princeton, New Jersey.

\bibitem[Clutton-Brock 1972]{Clutton-Brock:72}
Clutton-Brock, M. 1972,
\newblock Astrophys. Space. Sci., 16, 101.

\bibitem[Clutton-Brock 1973]{Clutton-Brock:73}
Clutton-Brock, M. 1973,
\newblock Astrophys. Space. Sci., 23, 55.

\bibitem[{Conselice} et~al. 2000]{Conselice.etal:00}
{Conselice}, C.~J., {Bershady}, M.~A., and {Jangren}, A. 2000,
\newblock \apj, 529, 886.

\bibitem[Courant and Hilbert 1953]{Courant.Hilbert:53}
Courant, R. and Hilbert, D. 1953,
\newblock {\em Methods of Mathematical Physics}, Vol.~1,
\newblock Interscience, New York.

\bibitem[Dahlquist and Bjork 1974]{Dahlquist.Bjork:74}
Dahlquist, G. and Bjork, A. 1974,
\newblock {\em Numerical Methods},
\newblock Prentice-Hall, Englewood Cliffs.

\bibitem[Hall 1981]{Hall:81}
Hall, P. 1981,
\newblock Ann. Stat., 9, 683.

\bibitem[Hernquist 1990]{Hernquist:90}
Hernquist, L. 1990,
\newblock ApJ, 356, 359.

\bibitem[Hernquist and Ostriker 1992]{Hernquist.Ostriker:92}
Hernquist, L. and Ostriker, J.~P. 1992,
\newblock ApJ, 386(2), 375.

\bibitem[Hjorth 1994]{Hjorth:94}
{Hjorth}, J. 1994,
\newblock \apj, 424, 106.

\bibitem[Ikeuchi et~al. 1974]{Ikeuchi.Nakamura.ea:74}
Ikeuchi, S., Nakamura, T., and Takahara, F. 1974,
\newblock Prog. Theor. Phys., 52.

\bibitem[{Julian} and {Toomre} 1966]{Julian.Toomre:66}
{Julian}, W.~H. and {Toomre}, A. 1966,
\newblock \apj, 146, 810.

\bibitem[Kalnajs 1976]{Kalnajs:76}
Kalnajs, A.~J. 1976,
\newblock ApJ, 205, 751.

\bibitem[Kalnajs 1977]{Kalnajs:77}
Kalnajs, A.~J. 1977,
\newblock ApJ, 212(3), 637.

\bibitem[Krall and Trivelpiece 1973]{Krall.Trivelpiece:73}
Krall, N.~A. and Trivelpiece, A.~W. 1973,
\newblock {\em Principles of Plasma Physics},
\newblock McGraw-Hill, New York.

\bibitem[Lin and Shu 1964]{Lin.Shu:64}
Lin, C.~C. and Shu, F. 1964,
\newblock ApJ, 140, 646.

\bibitem[Morse and Feshbach 1953]{Morse.Feshbach:53}
Morse, P.~M. and Feshbach, H. 1953,
\newblock {\em Methods of Theoretical Physics},
\newblock McGraw Hill, New York.

\bibitem[Pawula 1967]{Pawula:67}
Pawula, R.~F. 1967,
\newblock Phys. Rev., 162, 186.

\bibitem[Pruess and Fulton 1993]{Pruess.Fulton:93}
Pruess, S. and Fulton, C.~T. 1993,
\newblock ACM Trans. Math. Software, 63, 42.

\bibitem[Risken, 1989]{Risken:89}
Risken, H. 1989,
\newblock {\em The Fokker-Planck Equation},
\newblock Springer-Verlag.

\bibitem[{Robijn} and {Earn} 1996]{Robijn.Earn:96}
{Robijn}, F. H.~A. and {Earn}, D. J.~D. 1996,
\newblock MNRAS, 282, 1129.

\bibitem[{Saha} 1993]{Saha:93}
{Saha}, P. 1993,
\newblock MNRAS, 262, 1062.

\bibitem[{Shu} 1970a]{Shu:70a}
{Shu}, F.~H. 1970a,
\newblock \apj, 160, 89.

\bibitem[{Shu} 1970b]{Shu:70b}
{Shu}, F.~H. 1970b,
\newblock \apj, 160, 99.

\bibitem[Toomre 1969]{Toomre:69}
Toomre, A. 1969,
\newblock ApJ, 158, 899.

\bibitem[{Vesperini} and {Weinberg} 2000]{Vesperini.Weinberg:00}
{Vesperini}, E. and {Weinberg}, M.~D. 2000,
\newblock \apj, 534, 598.

\bibitem[Weinberg 1999]{Weinberg:99}
Weinberg, M.~D. 1999,
\newblock AJ, 117, 629.

\bibitem[{Weinberg} 2000a]{Weinberg:00b}
{Weinberg}, M.~D. 2000a,
\newblock {\em Noise-driven evolution in stellar systems: Theory, \rm
  submitted to MNRAS, \tt astro-ph/0007275}.

\bibitem[{Weinberg} 2000b]{Weinberg:00a}
{Weinberg}, M.~D. 2000b,
\newblock {\em Noise-driven evolution in stellar systems: A universal
  halo profile, \rm submited to MNRAS, \tt astro-ph/0007276}.

\end{susspbibliography}

\end{document}